\documentstyle[psfig,12pt,citesort]{article} 
\def \ov {\over}

\def \t {\theta} 
\def \p {\phi}

\newcounter{subequation}[equation]

\def\pa{\partial} 
 
\def\rt{\rightarrow} 
 
\newcommand{\be}{\begin{equation}} 
\newcommand{\ee}{\end{equation}} 
\newcommand{\eel}[1]{\label{#1}\end{equation}} 
\newcommand{\bea}{\begin{eqnarray}} 
\newcommand{\eea}{\end{eqnarray}} 
\newcommand{\eeal}[1]{\label{#1}\end{eqnarray}} 
\makeatletter 
\def\thesubequation{\theequation\@alph\c@subequation} 
\def\@subeqnnum{{\rm (\thesubequation)}} 
\def\slabel#1{\@bsphack\if@filesw {\let\thepage\relax 
   \xdef\@gtempa{\write\@auxout{\string 
      \newlabel{#1}{{\thesubequation}{\thepage}}}}}\@gtempa 
   \if@nobreak \ifvmode\nobreak\fi\fi\fi\@esphack} 
\def\subeqnarray{\stepcounter{equation} 
\let\@currentlabel=\theequation\global\c@subequation\@ne 
\global\@eqnswtrue 
\global\@eqcnt\z@\tabskip\@centering\let\\=\@subeqncr 
 
$$\halign to \displaywidth\bgroup\@eqnsel\hskip\@centering 
  $\displaystyle\tabskip\z@{##}$&\global\@eqcnt\@ne 
  \hskip 2\arraycolsep \hfil${##}$\hfil 
  &\global\@eqcnt\tw@ \hskip 2\arraycolsep 
  $\displaystyle\tabskip\z@{##}$\hfil 
   \tabskip\@centering&\llap{##}\tabskip\z@\cr} 
\def\endsubeqnarray{\@@subeqncr\egroup 
                     $$\global\@ignoretrue} 
\def\@subeqncr{{\ifnum0=`}\fi\@ifstar{\global\@eqpen\@M 
    \@ysubeqncr}{\global\@eqpen\interdisplaylinepenalty \@ysubeqncr}} 
\def\@ysubeqncr{\@ifnextchar [{\@xsubeqncr}{\@xsubeqncr[\z@]}} 
\def\@xsubeqncr[#1]{\ifnum0=`{\fi}\@@subeqncr 
   \noalign{\penalty\@eqpen\vskip\jot\vskip #1\relax}} 
\def\@@subeqncr{\let\@tempa\relax 
    \ifcase\@eqcnt \def\@tempa{& & &}\or \def\@tempa{& &} 
      \else \def\@tempa{&}\fi 
     \@tempa \if@eqnsw\@subeqnnum\refstepcounter{subequation}\fi 
     \global\@eqnswtrue\global\@eqcnt\z@\cr} 
\let\@ssubeqncr=\@subeqncr 
\@namedef{subeqnarray*}{\def\@subeqncr{\nonumber\@ssubeqncr}\subeqnarray} 
 
\@namedef{endsubeqnarray*}{\global\advance\c@equation\m@ne% 
                           \nonumber\endsubeqnarray} 
 
\makeatletter \@addtoreset{equation}{section} \makeatother 
\renewcommand{\theequation}{\thesection.\arabic{equation}} 
\catcode`\@=11 
 
\newcount\hour 
\newcount\minute 
\newtoks\amorpm \hour=\time\divide\hour by 60\minute=\time{\multiply\hour by 60 \global\advance\minute by-\hour} 
\edef\standardtime{{\ifnum\hour<12 \global\amorpm={am}% 
        \else\global\amorpm={pm}\advance\hour by-12 \fi 
        \ifnum\hour=0 \hour=12 \fi 
        \number\hour:\ifnum\minute<10 
        0\fi\number\minute\the\amorpm}} 
\edef\militarytime{\number\hour:\ifnum\minute<10 
0\fi\number\minute} 
 
\def\draftlabel#1{{\@bsphack\if@filesw {\let\thepage\relax 
   \xdef\@gtempa{\write\@auxout{\string 
      \newlabel{#1}{{\@currentlabel}{\thepage}}}}}\@gtempa 
   \if@nobreak \ifvmode\nobreak\fi\fi\fi\@esphack} 
        \gdef\@eqnlabel{#1}} 
\def\@eqnlabel{} 
\def\@vacuum{} 
\def\marginnote#1{} 
\def\draftmarginnote#1{\marginpar{\raggedright\scriptsize\tt#1}} 
\def\draft{ 
        \pagestyle{plain} 
        \overfullrule=2pt 
        \oddsidemargin -.5truein 
        \def\@oddhead{\sl \phantom{\today\quad\militarytime} \hfil 
        \smash{\Large\sl DRAFT} \hfil \today\quad\militarytime} 
        \let\@evenhead\@oddhead 
        \let\label=\draftlabel 
        \let\marginnote=\draftmarginnote 
        \def\ps@empty{\let\@mkboth\@gobbletwo 
        \def\@oddfoot{\hfil \smash{\Large\sl DRAFT} \hfil} 
        \let\@evenfoot\@oddhead} 
 
\def\@eqnnum{(\theequation)\rlap{\kern\marginparsep\tt\@eqnlabel}% 
        \global\let\@eqnlabel\@vacuum}  }

\renewcommand{\theequation}{\thesection.\arabic{equation}} 
\renewcommand{\thefootnote}{\fnsymbol{footnote}}

\def\appendix#1{ 
 \addtocounter{section}{-5} 
 \setcounter{equation}{0} 
  \renewcommand{\thesection}{\Alph{section}} 
  \section*{Appendix \thesection\protect\indent \parbox[t]{11.15cm} 
  {#1} } 
  \addcontentsline{toc}{section}{Appendix \thesection\ \ \ #1} 
  } 
 
\def \ov {\over}

\def \t {\theta} 
\def \p {\phi}

\def \LL{{\cal L}} 

\def\O{\Omega} 
 
\def\pd{\partial}

\def\m{\mu} 
 
\def\a{\alpha} 
\def\b{\beta}

\def\r{\rho}

\def\l{\lambda} 
 \def \t {\theta}

\def\p{\phi}

\def \foot{ \footnote} 
\def\be{\begin{equation}} 
\def\ee{\end{equation}}

\def \l {\lambda}

\def \m {\mu}

\def \rr {{\bar \rho}}

\def\l{\lambda}

\def\half{{1\over 2}}

\def\gym{g_{YM}} 
\def\gef{g_{eff}} 
\newcommand{\baq}{\begin{equation}\begin{array}{rcl}} 
\newcommand{\eaq}{\end{array}\end{equation}} 
\newcommand{\eaql}[1]{\end{array}\label{#1}\end{equation}} 
\newcommand{\beac}{\begin{equation}\begin{array}{rcl}} 
\newcommand{\eeacn}[1]{\end{array}\label{#1}\end{equation}} 
\newcommand{\ba}{\begin{array}} 
\newcommand{\ea}{\end{array}} 
\newcommand{\non}{\nonumber \\} 
 
\def\pg{Penrose-G\"uven limit} 
 
\date{} 
\begin{document} 
%\draft 
 
\begin{titlepage}

\hfill hep-th/0206033 
 
\hfill MCTP-02-26\\ 
 
\begin{center} 
\vskip 2.5 cm 
 
{\Large \bf Penrose Limits and  RG Flows} 
 
\vskip .7 cm

\vskip 1 cm 
 
{\large Eric Gimon${}^1$,  Leopoldo A. Pando Zayas${}^{1,2}$ 
and Jacob Sonnenschein${}^{1,3}$}\\ 
 
\end{center} 
 
\vskip .4cm \centerline{\it ${}^1$ School of Natural Sciences} 
\centerline{ \it Institute for Advanced Study} 
\centerline{\it Princeton, NJ 08540}

\vskip .4cm \centerline{\it ${}^2$ Michigan Center for Theoretical 
Physics} 
\centerline{ \it Randall Laboratory of Physics, The University of 
Michigan} 
\centerline{\it Ann Arbor, MI 48109-1120}

\vskip .4cm \centerline{\it ${}^3$ School of Physics and Astronomy} 
\centerline{ \it Beverly and Raymond Sackler Faculty of Exact Sciences} 
\centerline{ \it Tel Aviv University, Ramat Aviv, 69978, Israel}

\vskip 1.5 cm 
 
\begin{abstract} 
%We study the  Penrose limit of SUGRA backgrounds that are dual to 
%non-conformal gauge theories. 
%We show that the  quantum particle on this background is exactly solvable. 
%We write down explicitly two  Penrose limits at the IR fixed point 
%of the Pilch Warner solution. We analyze the corresponding gauge theory picture 
%and write down the operators which  are the duals of the  low lying string states. 
%The pp--wave limit of the large $N$  $D_p$ branes is written down.  In the far IR it is 
%characterized by negative mass$^2$. This phenomenon signals in the world sheet picture the 
%necessity to transform to another description $M_2$ from $D_2$ and $F_1$ from $D_1$ . 
 
The \pg\, simplifies a given supergravity solution into a pp-wave 
background. Aiming at clarifying its relation to renormalization 
group flow we study the \pg\, of supergravity backgrounds that 
are dual to non-conformal gauge theories. The resulting 
backgrounds fall in a class simple enough that the quantum 
particle  is exactly solvable. We propose a map between the 
effective time-dependent quantum mechanical problem and the RG 
flow in the gauge theory. As a testing ground we consider 
explicitly two Penrose limits of the infrared fixed point of the 
Pilch-Warner solution. We analyze the corresponding gauge theory 
picture and write down the operators which  are the duals of the 
low lying string states. We also address  RG flows of a different 
nature by considering the \pg\, of a stack of $N$ $D_p$ branes. 
We note that in the far IR (for $p<3$ )the limit generically has 
negative mass$^2$. This phenomenon signals, in the world sheet 
picture, the necessity to transform to another description. In 
this regard, we consider explicitly the cases of $M2$ from $D2$ 
and $F1$ from $D1$ .

\end{abstract} 
 
\end{titlepage}

\setcounter{page}{1} 
\renewcommand{\thefootnote}{\arabic{footnote}} 
\setcounter{footnote}{0} 
 
\def \N{{\cal N}} \def \ov {\over} 
 
\tableofcontents\vskip 0.5em\noindent\rule\textwidth{.1pt}\vskip 
0.5em 
 
%%%%%%%%%%%%%%%%%%%%%%%%%%%%%%%%%%%%%%%%%%%%%%%%%%%%%%%%%%%%%%%%% 
\section{Introduction} 
%%%%%%%%%%%%%%%%%%%%%%%%%%%%%%%%%%%%%%%%%%%%%%%%%%%%%%%%%%%%%%%% 
In spite of the remarkable successes of the gauge/gravity duality 
\cite{ads}, it has become clear that to penetrate to the core of the 
real world gauge dynamics, one will have to go beyond the SUGRA 
description and extract information from the related string theory. 
Superstring theories  on backgrounds like the  $AdS_5\times S^5$ 
are still far from being tractable. However, a  breakthrough in 
turning the string theory into a  solvable one, has been recently 
achieved~\cite{bmn} by taking the Penrose 
limit\cite{penrose,gueven} (see also \cite{tratado,figueroa})in 
the neighborhood of a suitably selected null geodesic. This led 
to an exactly solvable superstring theory \cite{metsaev,metse}. 
In the SYM side of the duality relation this  limit translates 
into a projection on operators with large R-charge $J$  such 
that  $g_{YM}^2N\over J^2$ is fixed.  So far most of the 
applications of the Penrose limit technique  were on SUGRA 
backgrounds that are the holographic duals of conformal field 
theories. The gauge interactions that we detect in nature are 
characterized by a scale and hence are not conformaly invariant. 
Thus, a natural step to take is to apply the \pg\, on SUGRA 
backgrounds associated with non-conformal gauge theories. 
 
Indeed, the goal of the present work is to analyze the pp--wave 
limit of SUGRA backgrounds that are dual to non-conformal field 
theories. Whereas in the context of the gauge/gravity duality the 
renormalization group flow was mapped into the variation of the 
SUGRA background fields, our main idea is to study the map 
between the  field theory RG flow and the world-sheet time 
evolution of the corresponding string theory.  We argue that the 
essential part of this map is in fact captured by the system of a 
quantum particle on the world-sheet(line) 
time-dependent pp--wave (tdppw) 
background. We show that this time-dependent quantum system can 
be {\em solved exactly}. 
 
The best situation to develop our intuition is the flow between 
two fixed points. An example of a property of the RG flow of the 
gauge theory that might be computed in the QM system is the 
mixing between operators.  Operators at the UV fixed point flow 
into the IR fixed point where they mix with several operators 
that carry their quantum numbers. In the dual QM system we can 
compute transition probabilities between initial state and final 
states of a time dependent QM harmonic oscillator.  This yields 
the mixing between gauge theory states with short characteristic 
scales (UV) and gauge theory state with long characteristic scales 
(IR) which point to a mixing between operators in the UV and the 
IR which we still do not quite understand.  The tdppw string 
theory generically differ from the conformal one in that it has 
additional non-trivial background fields like the dilaton, $NSNS$ 
$B_{\mu\nu}$ field and additional RR forms. We briefly discuss 
the effect of these fields.

We then demonstrate the general behavior of the tdppw duality in 
two classes of examples: (i) The Pilch Warner (PW) 10-d SUGRA 
solution \cite{pw} which is dual to the 
Leigh-Strassler flow \cite{strassler} of a UV ${\cal 
N}=4$ SYM theory into an IR ${\cal N}=1$  superconformal YM 
theory. (ii) The near horizon limit of large $N$ $D_p$ 
branes \cite{cobi}. In 
the former case we explicitly construct the Penrose limit of the 
IR fixed point of the PW solution. In fact we show that there are 
two possible natural limits that one can take. In the geometry 
side the two limits are associated with different choices of 
directions for the geodesic lines around which we zoom, whereas 
in the field theory side they are related to different sets of 
conformal operators that are the duals of the low lying string 
states. As opposed to the pp--wave limit of the ${\cal N}=4$ SYM 
here the string theory is characterized by non-trivial complex 
field-strength three form and hence a $B_{\mu\nu}$ term in the 
world action.  Correctly matching to the field theory dynamics 
will require the inclusion of the effects of this B-field on the 
world sheet Hamiltonian. 
 
 For the second class of examples, the SUGRA backgrounds of large $N$ 
$D_p$ branes, we derive the explicit metric, dilaton and $p+2$ 
forms expressed in the Poincar\'{e} coordinates. An interesting 
result of these limits is the appearance of negative $\mu^2$ 
terms.  We show that this is the generic behavior  apart from 
certain small region of the radial coordinate.  We discuss the 
meaning of these negative masses and show that the systems do not 
suffer from any pathological behavior. This is due to the fact 
that  in the far IR one has to transform into another description 
of the system which is characterized by positive  $\mu^2$ terms. 
In the case of $D_1$--brane in the region of large dilaton, an 
S-duality transformation render the system into that of large $N$ 
fundamental strings and for the case of $p=2$ the $D_2$ is 
replaced by an M2. In both cases the corresponding squared masses 
return to being positive.  Thus changing our description of the 
Hilbert space of states fixes any pathological runaway behavior. 
 
The analysis of Penrose limits of backgrounds dual to field 
theories that are not conformal was first addressed in \cite{pzs}, 
where the Schwarschild black hole in AdS, the small resolution of 
the conifold and the Klebanov-Tseytlin background were 
considered. Other papers that discussed related issues in pp-wave 
backgrounds \cite{holography}. The results of this paper were 
previously presented in \cite{presentations}. While preparing the 
manuscript for submission we received \cite{warner} which overlaps 
with section 4 of our work.

In section 2 we discuss the general structure of the bosonic 
string on world-sheet 
time dependent pp--wave (tdppw) background.  We start 
with the description of the geometry on the tdppw background, the 
additional forms and the dilaton and the relation between them 
via the SUGRA equation of motion.  We then analyze the quantum 
particle on this background. Since this system is in fact 
identical to a set of harmonic oscillators with time dependent 
frequencies, we review some of the techniques for solving such 
systems. In particular we write down the Schroedinger wave 
function and certain  transition probabilities. We show that the 
time-dependent frequencies associated with the tdppw limit have a 
form that enable us to have an {\em exact solution} to the QM 
system.  We then briefly discuss the bosonic string on this 
background.  Section 3 is devoted to a discussion of the RG flow 
of the gauge theory operators that associate with the tdppw 
limit. We review the operator/state duality of \cite{bmn} and 
discuss its alteration for the tdppw cases. We adress the issue 
of the Calan Symazic equation and its stringy analog. 
 
We then address our examples for the tdppw duality. We discussed 
our first example in section 4: the Penrose limits of the 
Pilch-Warner SUGRA solution that describes a RG flow from ${\cal 
N}=4$ to ${\cal N}=1$ SYM theories. We briefly review the the 
Pilch-Warner (PW ) solution.  The Penrose limit at the UV fixed 
point is reduced to that of the  ${\cal N}=4$ SYM discussed in 
\cite{bmn}.  We show that there are  two  different  Penrose 
limits at the  ${\cal N}=1$  fixed point which correspond to 
different geodesic lines.   We analyze the associated string 
states and the corresponding gauge field operators. We also write 
an expression of the masses of the  $AdS_5$ coordinates from the 
Penrose limit of the SUGRA flow in between the fixed points.  The 
second class of examples are the pp--wave limit of the near 
horizon limit of the large number of $D_p$ branes are discussed in 
section 5.  We perform the explicit limit in the Poincar'e 
coordinates and deduce the masses that appear in the string world 
sheet action.  We show that apart from a small region of the 
radial coordinate, the values of $\m_u^2$ are negative. The 
source of this behavior is the  non-constant dilaton. We discuss 
the meaning  of the negative world-sheet mass term for the 
propagating string.  We analyze separately the cases of $p=1$ 
where there is a flow from $D1$ to $F1$ string as well as the 
flow form the $D_2$ solution to the uplifted $M2$ background. The 
masses in the proper IR picture turn out to be positive.

%%%%%%%%%%%%%%%%%%%%%%% 
\section{ Strings on a time-dependent  pp--wave background: 
General structure  } 
\subsection {The geometry of the pp-wave limit of 
non-conformal backgrounds}

The general form of the pp--wave metric in the Brinkman 
coordinates is given by 
 \be 
 ds^2 = -4dx^+dx^-- \mu^2_{ij}(x^+)x^ix^j(dx^+)^2 + dx^i dx^i 
 \ee 
where $i,j=1...8$ This metric has an  isometry group that 
includes the shift of $x^-$ and  rotations in the eight 
dimensional space of $x^i$ which leave $\mu^2_{ij}$ invariant, 
for instance $SO(8)$ for $\mu^2_{ij}=\mu^2\delta_{ij}$. Isometry 
in a shift of $x^+$ occurs only provided that $\mu^2$ remains 
constant. 
 
Any  pp--wave metric can be  achieved by taking the Penrose limit 
of a certain parent metric, as we discuss in appendix A.  If the 
parent metric is conformaly invariant then the resulting 
$\mu^2_{ij}(x^+)$ is a constant. For instance, in the $AdS_5\times 
S^5$ parent metric expressed in global coordinates, the daughter 
Killing vector is $\pa_{x^+}=\pa_t+\pa_\psi$ where $\pa_t$ is 
associated with conformal transformation and $\pa_\psi$ with R 
symmetry. 
 
Non-conformal backgrounds expressed in terms of global coordinates 
are characterized by $G_{00}$ which is a function of time and 
hence shifting the global time $t$ is no longer a symmetry. 
 
The curvature of the pp--wave metric is characterized by one 
non-trivial component of the Ricci tensor and vanishing scalar 
curvature. $$ R_{++}=Tr[\mu_{ij}^2(x^+)]=\sum_i^8 \mu_i^2(x^+) 
\qquad R =0$$ where $\mu_i(x^+)$ are the eigenvalues of 
$\mu_{ij}(x^+)$.

In addition to the metric, the pp--wave background is 
characterized by a dilaton and certain forms. In the string frame 
the relation between the Ricci curvature, the dilaton and the five
form is given by 
 \be 
 R_{MN}= -2\nabla_M\nabla_N\phi + {1\over 
 96}e^{2\phi}[F_{M...}F_{N}^{...} -{1\over 10}G_{MN}F_{...}F^{...}] 
 \ee 
where $_{...}\ ^{...}$ denotes contracted indices. Since only $ 
R_{++}\neq 0 $ this reduces to 
 \be\label{rpp} 
 R_{++}= -2\pa_+\pa_+\phi + {1\over 96}e^{2\phi}[F_{+...}F_{+...}]. 
 \ee 
In section 4 we will discuss a background that include also a three
form. For that case one can easily amend  (\ref{rpp}).  
%%%%%%%%%%%%%%%%%%%%% 
\subsection{ The particle on the tdppw background } 
Prior to analyzing the  bosonic string on the tdppw background, 
we start with the simpler case of a particle propagating on this 
background. The system is that of a set of harmonic oscillators 
with time-dependent frequencies. The corresponding classical 
action takes the form 
 \be\label{particleaction} S_P= {1\over 2} 
\int d\tau  [ \half(  (\dot x^i)^2 - \mu^2_{i}(\tau)(x^2) ] 
 \ee 
where $\mu^2_{ij}$ has been  taken to be diagonal. The classical 
equations of motion 
 \be 
 \ddot x^i + \mu_i^2(\tau) x^i=0 
 \ee 
are  solved  by 
 \be 
x^i= a^iv^i + {a^i}^\dagger {v^i}^* 
 \ee 
where $v^i(t)$ has to obey 
 \be \label{v} 
\ddot v^i +\mu_i^2(\tau) v^i=0 
 \ee 
where there is no summation over the $i$ indices. 
The conjugate momenta are $p^i= a^i\dot v^i + {a^i}^\dagger \dot {v^i}^*$ 
and the Hamiltonian takes the form 
 \be 
H_P=\half[(p^i)^2+ \mu_i^2(\tau)(x^i)^2] 
 \ee 
 
We will now perform  the  canonical quantization of this  system 
using two different methods. In the first one we elevate the 
classical $a$ and $a^\dagger$ into 
  creation and annihilation operators 
$a$ and $a^\dagger$ 
 \be\label{ai} 
a^i(\tau) = i[{v^i}^*(\tau) p^i(\tau) -\dot {v^i}^*(\tau) x^i(\tau)]\qquad 
{a^i}^\dagger = -i[v^i(\tau) p^i(\tau) -\dot v^i(\tau) x^i(\tau)] 
 \ee 
To obey the standard quantization condition 
 $[a^i,{a^j}^\dagger]=\delta^{ij}$. 
one has to   impose the following further condition on $v^i$ 
 \be\label{further} 
[\dot {v^i}^*v^i - \dot v^i {v^i}^*] =-i 
 \ee 
The creation  and annihilation operators are ``time independent'' 
since 
 \be {da^i\over d\tau}= {\pa a^i\over \pa \tau} +i [H(\tau), a^i(\tau)] =0 
 \ee 
 Thus we can now define a Fock space of 
states 
 \be\label{Fock} a^i|0>=0; \qquad |n_i>={[{a^i}^\dagger]^n\over n!}|0> 
 \ee 
 
In terms of $a$  and $a^\dagger$ the Hamiltonian takes the form 
\be 
H_P= \half[((\dot v^i)^2 +\mu_i^2 (v^i)^2)(a^i)^2 + ((\dot {v^i}^*)^2 
+\mu_i^2 ({v^i}^*)^2)({a^i}^\dagger)^2 
+((|\dot v^i|)^2 +\mu_i^2 (|v^i|)^2)({a^i}^\dagger a^i + a^i{a^i}^\dagger) 
\ee 
For the special case of constant $\mu_i^2=w^2$ it is easy to check that 
$v^i(t)= {e^{i w^it}\over \sqrt {2w^i}}$ and then the Hamiltonian 
takes the usual form of  $H=\half w^i ({a^i}^\dagger a^i +  a^i {a^i}^\dagger )$ such that $H|n_i>=(\half+n_i)w^i$.

The  expectation values  of the  Hamiltonian at any time  for 
the states $|n> $are 
\be\label{Heigen} 
<n_i|H_P|n_i> = 
 (\half + n_i) [|\dot v^i(\tau)|^2 + \mu_i^2(\tau)|v^i(\tau)|^2] 
\ee 
Note however that these states do not diagonalize the Hamiltonian. 
We will see shortly that one can define instead of the Hamiltonian 
 another hermitian operator which is time independent and is diagonalized by 
(\ref{Fock}). Meanwile 
we  write the Schroedinger  wave function of the $n_i$ state\cite{kim} 
\be\label{WF} 
\Psi_{n_i}(x^i,\tau)=  \left[ 1\over 2\pi |v^i|^2\right]^{1/4} 
{1\over \sqrt{2^n_i n_i!}} \left[ {v^i\over {v^i}^*}\right ] 
H_{n_i} ({x_i \over \sqrt{2|v_i|^2}}) e^{i{\dot {v^i}^*\over {v^i}^*}x^2} 
\ee 
%%%%%%%%%%%%%%%%%% 
 
Consider the following invariant\cite{LewRie} 
\be 
I = \half\left [{1\over |v|^2}x^2 + ( |v|p - |\dot v| x)^2 \right ] 
\ee 
where $v(\tau)=|v|(\tau)e^{-i\varphi(\tau)}$  was  defined in(\ref{v}). 
To simplify the notation we discuss from here on a single oscillator. 
It is easy to check that $I$ is hermitian and that it is  an invariant, namely 
\be 
{dI\over d\tau}= {\pa I\over \pa \tau} +i [ H(\tau), I(\tau)]  =0 
\ee 
We construct a new  Fock space by using $\tilde a$ and $\tilde a^\dagger$ 
given by 
\be 
\tilde a = {1\over \sqrt{2}} \left [{1\over |v|} x + i( |v|p -|\dot v| x)\right ] 
\ee 
which can easily be related to  $a$, $a=\tilde a e^{i\varphi(\tau)}$. 
In terms of $\tilde a$ and $\tilde a ^\dagger$ the invariant takes the form 
\be 
I = \tilde a^\dagger \tilde a + \half 
\ee 
such that the state 
$ |\tilde n>={[{\tilde a}^\dagger]^n\over n!}|0>$ has an eigenvalue $I|\tilde n>= 
\lambda_n|\tilde n>= 
(n+1/2)|\tilde n>$ for the invariant 
$I$. Note that the invariant can also be expressed in the previous base since 
$I=a^\dagger a + 1/2$. 
We can now expand the general solution of the Schrodinger equation in terms of the 
new  base 
$ |\tilde n>$. 
\be 
\Psi = \sum_{\tilde n} c_{\tilde n}e^{i\alpha_{\tilde n}(\tau)}|\tilde n> 
\ee 
where $c_{\tilde n}$ are constants and 
 $\alpha_{\tilde n}(\tau) = -(\tilde n+1/2)\int^\tau {dt'\over |v(t')|^2}$. 
 
%%%%%%%%%%%%%%%%%%%%%%%%%%%%%%%%%%%%%%%%% 
Of a particular interest  for us will be the case where $\mu_i^2$ ``flows'' 
between two fixed points,  constant  $\mu^{(i)}$ at the initial time  and   constant 
$\mu^{(f)}$ at the final time. 
In that case there is a natural Fock space in both ends. 
To determine the transition between  the initial and final states 
one can use the method of Bogolyubov transformation. Instead 
we follow the precedure of using the invariant. 
 %%%%%%%%%%%%%%%%%%%%%%%%%%%%%%%%%%%%%%%%%%%%% 
Suppose now that  at $\tau=-\infty$  the Hamiltonian is characterized by a constant $\mu^{(i)}$ 
so that $|v(\tau=-\infty)|={1\over \sqrt{\mu^{(i)}}}$ and 
the system is in  its ground state $|0,i>$. At $\tau\rt \infty$ the corresponding ``frequency'' is 
$\mu^{(f)}$,  and the system is at a state $|\tilde n,f>$. At the final time 
$|v(\tau)|$ can be parametrized as follows 
$|v(\tau)|={1\over \sqrt{\mu^{(f)}}}[\cosh(\delta) + \sinh(\delta) 
\sin ( 2\mu^{(f)}\tau+\tilde\delta) ]^{1/2}$ 
 where  $\delta$ and 
$\tilde\delta$ are real. The transition probability  between 
these initial and final states takes the form\cite{LewRie} 
 \be\label{transition} 
P_{0n}= 
{(2n)!\over2^{2n} (n!)^2 }\left [{\cosh(\delta) -1\over 
(\cosh(\delta) +1)}\right ]^{n/2} \left [ {2\over (\cosh(\delta) 
+1)}\right ]^{1/2} 
 \ee 
 for even $n$ and it vanishes for odd $n$. 
In the sudden approximation the transition probability takes the 
form 
 \be\label{sudden} 
 P_{0n}= {(2n)!\over2^{2n} (n!)^2 } \left 
[{ (\mu^{(f)}-\mu^{(i)})\over(\mu^{(f)}+ \mu^{(i)}  }\right]^{n/2} 
{2( \mu^{(f)} \mu^{(i)})^{1/2}\over \mu^{(i)}+ \mu^{(f)}} 
 \ee 
 
In certain cases it is useful to use the path-integral approach\cite{KhaLaw}. In this case 
the evolution of the wave function is expressed in terms of a Kernel $K(x',\tau';x,\tau)$ such 
that $\Psi(x,\tau)=\int dx'K(x',\tau';x,\tau)$. The Kernel is determined by the path integral 
$K(x',\tau';x,\tau)=\int D(\tilde x)e^{i\int_{\tau}^{\tau'}L d\tau''}$. 
An expression for the Kernel was written down in \cite{KhaLaw} and it takes the following form 
 \bea 
K(x',\tau';x,\tau) &=& \left [ {\sqrt{\dot\varphi'\dot \varphi}\over 2\pi i 
\sin( 
\varphi'-\varphi)} \right ]^{1/2} e^{i/2({|\dot v|'\over |v|'}{x'}^2-{|\dot v|\over |v|}x^2)}\non\\ 
&\times& e^{\left ({i\over 2 \sin(\varphi'-\varphi)}[\dot\varphi'{x'}^2+\dot\varphi{x}^2) 
\cos(\varphi'-\varphi)- 2\sqrt{\dot\varphi'\dot\varphi}{x'}x]\right )} 
 \eea 
where $'$ stands for a variable evaluated at $\tau'$ and $\dot A\equiv \pa_\tau A$. 
It is easy to check that for constant $\mu$ where $|v|={1\over \sqrt{\mu}}$  and $\varphi=w\tau$ the Kernel reduces to that of an ordinary harmonic oscillator. 
 
We can now apply these general considerations of a time dependent quantum 
 Harmonic oscillator 
to the particular  quantum mechanics on a tdppw backgrounds. 
It was shown in \cite{tratado} ( see also appendix A) that a  background with diagonal $\mu_{ij}^2$, 
the eigenvalues of the mass matrix  take  the following form 
 \be 
 \label{mudef} 
 \mu_i^2(\tau)=- {\ddot A_i \over A_i} 
 \ee 
where $A_i$ is given in terms of the metric. For instance in 
appendix A we have $A_i=\sqrt{g_{ii}}$ for $\mu_{x^i}$, 
 and $A_i=\sqrt{E^2 g_{\psi\psi}-\mu^2 
g_{tt}}$ for $\mu_\phi$. 
 It is easy to see that $v_i(\tau)=A_i(\tau)$ is a 
solution of the defining equation (\ref{v}). However since $v_i$ 
is in general complex and $A_i$ is real, it is clear that 
$v_i(\tau)=A_i(\tau)$  is not the most 
 general solution. It is easy to show that the latter takes the form 
 \be\label{vi} 
 v^i(\tau) = A_i(\tau)(c^i_0 + c^i_1\int {d\tau\over A_i^2}) 
 \qquad ({c^i}^*_0c^i_1-c^i_0{c^i}^*_1) = i 
 \ee 
where $c^i_0$ and $c^i_1$ are two time-independent  complex 
coefficients and  the condition they have to fulfill follows 
from  (\ref{further}). We would like to emphasize that the 
relation (\ref {ai}) implies that we have {\em an exact solution } 
of the quantum mechanical particle in the  tdppw backgrounds 
 we are considering. 
 It is now straightforward to deduce the 
eigenvalues of the Hamiltonian (\ref{Heigen}), 
 the Schroedinger wave function (\ref{WF}) and the transition probabilities 
 (\ref{transition}) just by substituting (\ref{ai}) for $v(\tau)$. 
 
%In section ... we will be interested in  SUGRA 
%backgrounds that are dual to field theories 
%that flow from an UV fixed point to 
%an IR fixed point. {\bf add fig} 
%This phenomenon translates into $H_P$  that have  fixed points as 
%$\tau\rt \pm\infty$ . 
 
%Note that we are using here a Heisenberg representation for which the states are time independent. 
%However, the definition of the GS is time dependent. Thus, the formalism specified here can be used only for slowly varying $\mu_i(\tau)$. 
 
The rate of change of the eigenvalues of the Hamiltonian in the 
basis (\ref{Fock}) 
 \be 
 \pa_\tau <n_i|H|n_i> = (n_i+1/2)\pa_\tau 
\mu_i^2(\tau)|v(\tau)|^2\sim [ (c_0 + c_1\int {d\tau\over A_i^2}]^2 
(\pa_\tau^3  A_i A_i - \pa_\tau^2 A_i \pa_\tau A_i ) 
 \ee 
 It is thus clear that the Hamiltonian has fixed points at 
 \be\label{fp} 
\pa_\tau^3 A_i A_i - \pa_\tau^2 A_i \pa_\tau A_i  =0 
 \ee 
This is clearly the same as demanding that $\mu$ be constant in 
eq. (\ref{mudef}). 
 
 We will be interested in cases where there 
is a fixed point at each end of the 
flow, namely, when (\ref{fp}) is obeyed for $A_i(\pm\infty)$. 
%A QM system that evolves    with   $\mu_i(\tau)$ of the form of fig.. 
%is characterized by the two fixed points values of $\mu_i$, namely 
%$\mu_i(-\infty)\equiv \mu^{(i)} $ and $\mu_i(\infty)\equiv \mu^{(f)} $ 
%Any state of the Fock space associated with $\mu^{(i)} $ will evolve into 
%an mixture of final states associated with   $\mu^{(f)} $ 
%Using the vawe fucntions of (\ref{wf}), it is easy to see for instance 
%that the inital and final ground state have the following overlap. 
%\bea 
%<\Psi_0(x_i, t&=&-\infty)|\Psi_0(x_i, t=-\infty)> \\ 
%&=& \int dx_i [  ]^{1/4}e^{-1/2 \mu^{(i)}  x_i^2} 
%[  ]^{1/4}e^{-1/2\mu^{(i)}x_i^2} = [\sqrt{\mu^{(i)} \mu^{(f)}\over 
%1/2(\mu^{(i)} +\mu^{(f)})}] 
%\eea 
%namely the square root of the ratio of the geometric average to the arithemtic 
%average of the frequencies at the two end points. 
%More generally  the mixture of an initial $n_i$ state to a $m_j$ one has an overlap of ... 

 %%%%%%%%%%%%%%%%%%%%%%%%%%%%%%%%%%%%%%%%%%%%%%%%& 

%%%%%%%%%%%%%%%%%%%%%%%%%%%%%%%%%%%%%%%%%%%%%%%%%%%%%%%%%%%%%%%%%%%%% 
 
\subsection{ The bosonic tdppw  string } 
 
%begin{itemize} 
The discussion of the particle in the previous subsection applies 
also to the zero modes of the  corresponding bosonic string on 
tdppw background.  In fact, as will be shown below, in the dual 
gauge theory picture the operators that correspond to the ground 
state and to the zero mode of the string are the basic building 
blocks of for the operators which are dual to the string states, 
and hence the quantum mechanical system captures the essence of 
the state/operator duality.  In this subsection we add several 
comments about the full bosonic string theory.  For the case 
where $\mu^2_{ij}(x^+)$ is a constant matrix the string theory 
has been analyzed in \cite{metsaev,metse}.  Here  we  focus on the 
properties that are directly  related to non-constant masses  and 
to additional fields  of the background.

The action of the  bosonic part of the superstring theory defined on a 
ppw  background is given by 
 
 \bea  S &=& {1\over 2\pi \alpha'}\int d^2 \sigma 
\sqrt{h}[h^{\alpha\beta}(-4\pa_\alpha x^+ \pa_\beta x^- - 
\mu^2_{ij}(x^+)x^ix^j \pa_\alpha x^+ \pa_\beta x^+ + \pa_\alpha  x^i 
\pa_\beta x^i ) +\alpha' R^{(2)}\phi(x^+)] \non\\  &+& {1\over 2\pi 
\alpha'}\int d^2 \sigma \epsilon^{\alpha\beta}( B_{ij}(x^+) \pa_\alpha 
x^i \pa_\beta x^i + B_{i+}(x^j) \pa_\alpha x^i \pa_\beta x^+ ) \eea 
 
For the case where the dilaton is a constant, the last term  in 
the first line is  just a constant times the Euler number of the 
world sheet and can  be ignored. In general the dilaton term is 
multiplied by  $\alpha'$ and hence it is a quantum correction 
that does not directly affect the equations of motion.  The 
$NS$--$NS$ $B_{\mu\nu}$ two form term will be relevant in  section 
4. In fact using a gauge transformation the  $B_{\mu\nu}$ term 
transforms in such a way that it contains only a $B_{ij}(x^+)$ 
component. This is in accordance with the  general form of  the 
pp--wave limit where any field strength has the form $F_{p+1}= 
dx^+ \wedge \pa_{x^+}A_p(x^+)$\, \cite{tratado}.  However as will 
be discussed in section 4. the gauge transformation may yield a 
non-trivial surface term. 
 
We now implement the light-cone gauge fixing. In \cite{horowitz} it 
was shown that the ppw background is the only curved space-time 
for which the gauge can be implemented  as in the flat case namely 
$\sqrt{h}h^{\alpha\beta}=\eta^{\alpha\beta}$ with $h^{00}=-1$ . In 
addition, since $x^+$ obeys the equation $\Box x^+=0$ due to the 
equation of motion for $x^-$, we fix the residual symmetry by 
setting $x^+(\tau,\sigma)=\tau$.  The gauged fixed action (for a 
constant dilaton)  thus takes the form 
 \be\label{action} 
S_B= {1\over 2\pi \alpha'}\int d\tau \int_0^{2\pi\alpha'p^+} 
d\sigma [ (  (\dot x^i)^2 - ({x^i}')^2 - 
\mu^2_{ij}(\tau)x^ix^j + B_{ij}(x^+) 
( \dot x^i{x^j}'- \dot x^j{x^i}')   +  B_{i+}(x^j){x^i}' ] 
 \ee 
where $'$ stands for derivative with respect to $\sigma$. 
For the case of vanishing $B_{\mu\nu}$ terms we have 
a  1+1 dimensional action of eight 
scalar fields with a time-dependent mass term. Note also that if one reduce the world-sheet into a world-line 
going from the string model to a particle model the $B_{\mu\nu}$ terms do not affect the QM picture so the 
discussion of the previous subsection remains valid.

The corresponding equations of motion take the form 
 \be 
 \ddot x^i -{x^i}'' +\mu_{ij}(\tau)x^j  -H_{+ij}(\tau) {x^j}' =0 
 \ee 
 where $H_{+ij}=\pa_{x^+}B_{ij}$ 
%Note that for the zero mode there is no  contribution from  the $B_{ij}$ term and hence the 
%analysis of the previous subsection applies also for this case. 
Unlike the quantum mechanical problem, even with a vanishing $ H_{+ij}$ we cannot write down 
a universal solution 
to the equations of motion. If we substitute $\mu_i^2(\tau)=- {\ddot A_i \over A_i}$ and assuming a dependence 
on $\sigma$ of the form $e^{ik_n\sigma}$ we get an equation of the form 
$ \ddot x^i +(k_n^2-{\ddot A_i \over A_i})x^i=$ which does not see to admit a universal solution. 
 
The bosonic part of the world sheet Hamiltonian $H$ associated 
with the action \ref{action}  is given by 
 
 \be 
 H=   \int_0^{2\pi\alpha'p^+} d\sigma [(p_i-B_{ij}(\tau){x^j}')^2 + ({x^i}')^2 
+\sum_i\mu^2_i(\tau){x^i}^2 ] 
 \ee 
 
Obviously this Hamiltonian is   not conserved  due to the $\tau$ dependence 
of the  mass term and the $B_{ij}$ term. 
 We quantize the system following similar steps to those taken for 
the the particle. 
%s zero stDefine now creation and annihilation operators 

%The eigenvalue of the  normal ordered hamiltonian  in the $|n^i>$ 
%is given by 
% \be 
% <n^i|H-H_0|n^i> = \sqrt{ n^2 + \alpha' p^+ [|\dot v^i(\tau)|^2 + \mu_i^2(\tau)|v^i(\tau)|^2]^2} 
% \ee 
% 
%This can be expressed in terms of $A^i$ by substituting  (\ref{ai}). 

%%%%%%%%%%%%%%%%%%%%%%%%%%%%%%%%%%%%%%%%%%%%%%%%%%%%%%% 
\section{ The RG flow  in the dual field theories} 
%%%%%%%%%%%%%%%%%%%%%%%%%%%%%%%%%%%%%%%%%%%%%%%%%%%%%%% 
 
  In \cite{bmn} the field theory duals of the pp--wave string states were 
identified with operators of large $J$. In particular the vacuum 
state $|0,p^+>$ is identified with the state dual to the operator 
${1\over \sqrt{J}N^{J/2}}Tr[Z^J]$ integrated over $S^3$ and an 
excited string state ${a^i}^\dagger_n{a^j}^\dagger_{-n}|0>$ with a 
state dual to 
\begin{equation} 
O_n={1\over \sqrt{JN^{J+2}}}\sum_{l=0}^J e^{2\pi i 
ln/J}Tr[\phi^iZ^{l-J}\phi^j Z^l], 
\end{equation} 
also integrated over $S^3$.  In fact, as was mentioned above, the 
basic building blocks are the GS and the first excited states with 
$n=0$.  It was further shown that the perturbative calculation 
with ${\lambda n^2\over J^2}$ as the  coupling constant 
reproduced the string theory spectrum. The world-sheet energy 
eigenvalues $w_n=\sqrt{1+{\lambda n^2\over J^2}}$ are mapped into 
the values of $(\Delta -J)_n$ of the corresponding operators. 
 
Throughout this work, the conformal nature of ${\cal N}=4$ SYM 
was explicitly used to map from the field theory on $R^{1,3}$ to 
a theory on $R\times S^3$ where the operator/state correspondence 
makes sense. To be more precise, suppose we start with a state 
$|\Psi(t,\vec{x})>$ in a conformal theory on ${R}^{1,3}$. We can 
use a conformal transformation to turn this into a state 
$|\Psi(t,\vec{\varphi})>$ on ${R}\times S^3$.  After a Euclidean 
continuation, one can see that these states correspond to placing 
and operator $\Phi$ at the origin of ${R}^4$ and then using 
radial quantization to get $|\Psi(\lambda,\vec{\varphi})>$.

A natural question to ask is what of the structure of the BMN 
limit also holds for non-conformal cases and to what extent can 
the behavior of strings on a tdppw background teach us about the 
properties of the RG flow of the gauge theory dual to the tdppw 
background's parent SUGRA solution.  States in the tdppw 
correspond to states in the SUGRA solution localized around a 
geodesic in the Poincar\'{e} patch going from large values of $u$ 
(UV) to smaller values of $u$ as $t\to\pm\infty$.  In the dual 
gauge theory we will start with a state, $|\Psi(t=0,\vec{x})>$, 
with short characteristic scales (UV) and it will evolve with 
time into a state,$|\Psi(t\gg 0,\vec{x})>$, with long 
characteristic scales(IR).  Suppose we are dealing with a gauge 
theory which is conformal in the IR and the UV.  Then near $t=0$ 
and $t\to\pm\infty$ we can use a separate state/operator map to 
get operators $\Phi_{UV}$ and $\Phi_{IR}$ respectively.  The time 
evolution of $|\Psi(t,\vec{x})>$ which is described by strings on 
the tdppw background should correspond to some kind of RG-flow on 
the operator $\Phi$. 
 
%Using the AdS/CFT duality, this implies that the operators 
%defined on the  boundary of the Poincar\'{e} patch are mapped 
%into states of a system described by the global coordinates of 
%the $AdS_5$. We can further perform a KK reduction on the $S^3$. 
%Now since the system is conformal invariant the integration over 
%the $S^3$ can be done at any radius of the $S^3$. Thus the 
%correspondence can be summarized as $\Phi^i\equiv \int_{S^3}|\Phi 
%(S^3)> \leftrightarrow {a^i}^\dagger|0>$ where $\Phi^i$ is a 
%primary operator of the CFT and ${a^i}^\dagger|0>$ is a state of 
%the string theory (or of the QM particle theory for the zero 
%mode). 
%For a non-conformal field theory there is no invariance under 
%dilatation of the radial direction so that the  map takes the 
%form of 
%$\Phi^i(\rho)\equiv \int_{S^3_\rho}|\Phi^i(S^3_\rho)> 
%\leftrightarrow {a^i}(\tau)^\dagger|0>$ 
 
Going back briefly to the ${\cal N}=4$ case in \cite{bmn}, it was 
shown that an eigenvalue of the world sheet Hamiltonian 
$H_{ws}\equiv 2P^-$ for a given state translates in the CFT 
picture to $\Delta - J$ of the corresponding operator, where 
$\Delta$ is the conformal dimension and $J$ is a global symmetry 
charge.  In the non-conformal case the analog of the 
time-dependence of the string Hamiltonian is the evolution of the 
gauge theory states in a scale dependent theory.  Near conformal 
points, this is just scale dependence for conformal dimensions. 
Strictly speaking it does not make sense to discuss the 
eigenvalues of a time dependent Hamiltonian just as it is not 
meaningful to discuss conformal dimension in a non-conformal 
invariant system. In time-dependent QM, what replaces a 
description in terms of eigenstates of the Hamiltonian is the 
evolution of the wave-function determined by the Schroedinger 
operator $i\partial_{\tau} -H(\tau)$.  In non-conformal field 
theories the RG flow is described by the Callan Symanzic (CS) 
equation which reads $({\cal M}\partial_{{\cal M}} + 
\beta\partial_g + n\gamma)G_n(x_1,...x_n,{\cal M})$  where $ 
{\cal M}$ is the scale of the system (Euclidean time), $\beta$ is 
the $\beta$ function of the coupling $g$, $\gamma$ is the 
anomalous dimension of the operator and $G_n$ is an n-point 
function which depends on the locations of the operators and the 
scale.  It is not completely clear to us how the RG flow for 
operators is related to the time evolution of the corresponding 
gauge theory states, but the connection is tantalizing. 
 
%If we integrate $G_n$ over the locations of the operators, the 
%product of the operators maps into 
%$<0|a^{n/2}(\tau){a^\dagger}^{n/2}(\tau)|0>$ in the string (or 
%QM) system.  It is thus tempting to conjecture that the analog of 
%the CS  equation is the equation achieved by applying the 
%Schroedinger operator on  ${a^\dagger}^n(\tau)|0>$. In the 
%conformal points the $\beta$ function part of the CS equation 
%vanishes leaving only the derivative with respect to the scale 
%and the anomalous  dimension term, which are the duals of 
%$i\partial_{\tau}$ and $H_{ws}+J-\Delta_0$ where $\Delta_0$ is 
%the naive dimension. Recall also that for $N=1$ supersymmetric 
%gauge theories the exact $\beta$ is given in terms of some group 
%theory factors and the anomalous dimensions\cite{SVZ} so that it 
%is plausible that the $H_{ws}$ of the tdppw string models encodes 
%also the information about the $\beta$ function. The precise 
%relation between the CS equation and its stringy counterpart 
%deserves further investigation. 
 
   The flow that will capture great part of our attention is between 
a fixed point in the UV and a fixed point in the IR, where the 
state/operator map can be well defined at both ends. An example of 
this nature will be discussed in details in section 4.  A 
schematic description of the evolution of the conformal dimension 
is drawn in fig. \ref{fig:mufig}. An operator that flows from the UV 
fixed point will mix with a set of operators in the IR that carry 
the same quantum numbers. The transition probabilities between 
the initial operator and the final ones maps into the transition 
probability of an initial state of the QM system described in the 
previous section to mix with certain final states. 
 
    Flows in gauge dynamical systems can take a variety of forms. In 
ordinary QCD there is a flow from asymptotic freedom in the UV to 
a mass gap in the IR.  There is a state/operator map in the UV, 
and we could hope that a pp--wave--like procedure might show how 
states created by higher spin operators in the UV mix with 
infrared states like baryons.

\section{The \pg \, of the Pilch-Warner solution} 
 
%%%%%%%%%%%%%%%%%%%%%%%%%%%%%%%%%%%%%%%%%%%%%%%%%%%%%%%%%%%%%%%%%%%%%%%%%%%%%%% 
 
The Pilch-Warner solution computed in ref.\cite{pw} is one of the 
premier examples of the RG flow in the AdS/CFT context. It is the 
full ten dimensional supergravity uplift of the FGPW solution 
\cite{fgpw}
describing the gravity dual of ${\cal N}=4$ supersymmetric 
$SU(N)$ Yang--Mills theory mass deformed to become the 
Leigh-Strassler ${\cal N}=1$ SCFT in the IR\cite{strassler}.

\subsection{The Pilch-Warner solution-a brief review} 
 
\label{tendee} The ten dimensional solutions computed in 
ref.\cite{pw} may be written as~\foot{We borrow liberally from 
the excellent review in \cite{cvj}, although we keep the 
normalization for the $\sigma$'s used in ref.\cite{pw}}: 
 
\begin{equation} 
ds^2_{10}=\Omega^2 ds^2_{1,4}+ds^2_{5}\ , 
 \label{fullmetric} 
\end{equation} 
 
for the Einstein metric, where 
 
\begin{equation} 
ds^2_{1,4}=e^{2A(r)}\left(-dt^2+dx_1^2+dx_2^2+dx_3^2\right)+dr^2\, 
\label{littlemetric} 
\end{equation} 
and 
\begin{eqnarray} 
ds_5^2&=&{L^2}{\Omega^2\over\rr^2\cosh^2\chi}\left[{d\theta^2} 
+\rr^6\cos^2\theta\left({\cosh\chi\over 
    {4\bar X}_2}\sigma_3^2+{\sigma_1^2+\sigma_2^2\over 
{4\bar X}_1}\right)\right.\nonumber\\ 
&&\left.+{{\bar X}_2\cosh\chi\sin^2\theta\over {\bar X}^2_1} 
\left({d\psi} +{\rr^{6}\sinh\chi \tanh\chi \cos^2\theta\over 
{2\bar X}_2} \sigma_3\right)^2\right]\ , 
 \label{bigmetric} 
\end{eqnarray} 
 
with 
\begin{eqnarray} 
\Omega^2&=&{{\bar X}_1^{1/2}\cosh\chi\over \rr}\nonumber\\ 
{\bar X}_1&=&\cos^2\theta+\rr^6\sin^2\theta\nonumber\\ 
{\bar X}_2&=&{\rm 
sech}\chi\cos^2\theta+\rr^6\cosh\chi\sin^2\theta\ . 
 \label{warp} 
\end{eqnarray} 
 
The $\sigma_i$ are the standard $SU(2)$ left--invariant forms, 
the sum of the squares of which give the standard metric on a 
round three--sphere of radius 2. They are normalized such that 
$d\sigma_i={1\over 2}\,\epsilon_{ijk}\sigma_j\wedge\sigma_k$.  We 
use the coordinates on the $S^3$ as $(\phi_1,\theta_1,\psi_1)$ and 
we will us the explicit form for the $\sigma^i$'s: 
 \be 
   \sigma_1 + i \sigma_2 = e^{i\phi_1}(d\theta_1 - i \sin\theta_1 
   d\psi_1),\qquad \sigma_3 = d\phi_1 + \cos\theta_1 d\psi_1. 
 \ee 
The functions $\rr(r)$ and $\chi(r)$ which appear here in the 
ten dimensional metric are a one--parameter family of functions 
which depend on the size of the initial mass deformation. 
 
At $r\to\infty$, the UV, the various functions in the solution 
have the following asymptotic values: 
\begin{equation} 
\rr(r)\to 1\ ,\,\, \chi(r)\to 0\ ,\,\, A(r)\to {r\over L}\ . 
\label{asymptotesUV} 
\end{equation}  This gives standard AdS$_5\times S^5$ with radius $L$. 
At $r\to\ 0$, the IR, we get new asymptotic values: 
\begin{equation} 
\rr(r)\to 2^{1\over 6}\ ,\,\, \sinh\chi(r)\to 3^{-{1\over 2}}\ 
,\,\, A(r)\to {r\over \tilde{L}}\ . \label{asymptotesIR} 
\end{equation} 
(with $\rr'=\chi'=A''=0$.  This yields AdS$_5$ with a new radius 
$\tilde{L} = 3 2^{-{5\over 3}} L$ times a deformed $S^5$ (see 
\cite{pw0}). There is no known exact solution for the particular 
equations which give $\rr(r),\chi(r)$ and $A(r)$ throughout the 
flow. 
 
%%%%%%%%%%%%%%%%%%%%%%%%%%%%%%%%%%%%%%%%%%%%%%%%%%%%%%%%%%%%%%%%%%%%%%%%%%%%%% 
\subsection{Penrose Limit of the Pilch-Warner solution in the IR} 
%%%%%%%%%%%%%%%%%%%%%%%%%%%%%%%%%%%%%%%%%%%%%%%%%%%%%%%%%%%%%%%%%%%%%%%%%%%%%% 
    In the IR limit the Pilch-Warner solution has three simple Penrose limits 
that we can identify. In the first limit we pick a trajectory on 
$S^5$ near $\theta={\pi \over 2}$ and along the $-\,\psi$ 
direction (i.e., $\psi = - \textrm{const.}\,\cdot\, x^+ + 
\ldots$), expanding to quadratic order in $\theta'$: $\theta={\pi 
\over 2} - \theta'$. The PW solution then takes the form: 
\begin{eqnarray} 
ds_{10}^2&=& \Omega^2{9L^2\over 2^{10/3}}\bigg[-\cosh^2\r d\tau^2 
+ d\r^2 +\sinh^2\r \,\, d\Omega_3^2\bigg] + L^2 ds_5^2, \nonumber \\ 
ds_{5}^2&=&{\sqrt{3}\over 2}\bigg[\,d\theta'^2+{\theta'^2\over 4} 
(d\theta_1^2+d\phi_1^2+d\psi_1^2 + 2\cos\theta_1 d\phi_1 d\psi_1) 
+ ({2\over 3}-{7\over 12}\,\theta'^2)d\psi^2 \nonumber \\ 
&&\qquad\quad +{\theta'^2\over 6} (d\phi_1 + \cos\theta_1 d\psi_1) 
 d\psi\bigg], \nonumber \\ 
\Omega^2&=& {2^{4/3}\over \sqrt{3}}-{\theta'^2\over 
2^{2/3}\sqrt{3}}. 
\end{eqnarray} 
If we now make the following coordinate transformation (the shift 
in in $\phi_1$ is not a change in the geodesic, but an 
$x^+$--dependent rotation of its transverse space): 
\begin{equation} 
\label{var1} 
 x^+ = {\sqrt{3} \over 4\mu}(\tau - {2\over 3}\,\psi), 
 \; x^- = \mu {L^2\over 2}({3\over 2}\,\tau + \psi), 
 \; \hat{\phi_1} = \phi_1 + {1\over 3}\,\psi, 
 \; \theta'^2 = { 2\,y^2 \over \sqrt{3} L^2}, 
 \; \rho^2 = {4\,r^2 \over 3\sqrt{3} L^2} 
\end{equation} 
we get the final form for the Penrose limit of the Pilch-Warner 
background \footnote{The expression for the ${\cal A}_{(2)}$ corrects a typo in reference 
\cite{pw}. We are thankful to K. Pilch for confirming this typo.}: 
 \begin{eqnarray} 
 \label{pimetric} 
 ds^2_{PP} &=& -4dx^+dx^- -{4\over 3}\mu [\vec r^2 
+ \vec y^2](dx^+)^2 + (d\vec r)^2 +( d\vec y)^2 + 
\emph{O}\,(L^{-1}) \nonumber \\ 
 {\cal A}_{(2)} &=&B_{NS-NS}+iB_{RR} \nonumber \\ 
 &=& - {\sqrt{3}\over 2}\,e^{i {2\over\sqrt{3}} \mu 
 x^+} d\bar{z}_1\wedge d\bar{z}_2 + {1\over \sqrt{6}}\,d\bigg[ 
e^{i {2\over\sqrt{3}} \mu x^+} (\bar{z}_1\wedge d\bar{z}_2 - 
\bar{z}_2\wedge d\bar{z}_1)\bigg] 
 \end{eqnarray} 
with $z_1 = y_1 + i y_2, z_2= y_3 + i y_4$ (the total derivative 
in ${\cal A}_{(2)}$ is important because it does not appear in 
the UV) and 
 \begin{eqnarray} 
 p_+ &=& \partial_{x^+} 
     = {2\over \sqrt{3}} \mu\,( \partial_{\tau} 
     - {3\over 2} \,\partial_{\psi} 
     + {1\over 2} \,\partial_{\psi_1}) 
 \nonumber \\ 
 p_- &=& \partial_{x^-} 
     = \;\,{2\over 3\mu}\;(\partial_{\tau} 
     + {3\over 2} \,\partial_{\psi} 
     - {1\over 2} \,\partial_{\psi_1}). 
 \end{eqnarray} 
 
 The second and third limit correspond to boosting along a path in 
$S^5$ corresponding to $\theta = 0,\; \theta_1= 0$ or $\pi$. These 
two limits are pretty much identical except for a change in the 
sign of $\psi_1$. Expanding the PW solution to quadratic order in 
$\theta$ and $\theta_1$ we get: 
\begin{eqnarray} 
ds_{10}^2&=& \Omega^2{9L^2\over 2^{10/3}}\bigg[-\cosh^2\r d\tau^2 
 + d\r^2 +\sinh^2\r \,\, d\Omega_3^2\bigg] + L^2 ds_5^2, \nonumber \\ 
ds_{5}^2&=&{\sqrt{6}\over 4}( 1 + {1\over 2} \theta^2) 
 \bigg[\,d\theta^2+ {2\over 3}(1 - {8\over 3}\theta^2) 
 (d\phi_1^2+d\psi_1^2 - \theta_1^2 d\phi_1\psi_1 - \theta_1^2 
 d\psi_1d\psi_1) \nonumber \\ 
 &&\quad + {1\over 2}(d\theta_1^2 + \theta_1^2\, d\psi_1^2) 
 +\theta^2 \bigg(d\psi + {1\over 3}(d\phi_1 + d\psi_1)\bigg)\bigg], \nonumber \\ 
\Omega^2&=& {2^{5/6}\over \sqrt{3}}( 1 + {1\over 2} \theta^2). 
\end{eqnarray} 
Now the following coordinate transformation is appropriate: 
\begin{eqnarray} 
 && x^+ = {\sqrt{6} \over 8\mu}(\tau + {2\over 3}\,(\phi_1 + \psi_1)), 
 \; x^- = \mu {L^2\over 2}({3\over 2}\,\tau - (\phi_1 + \psi_1)), 
 \; \hat{\psi_1} = {2\over 3}\,\phi_1 - {1\over 3}\,\psi_1, \nonumber \\ 
 && \hat{\psi}   = \psi + {1\over 3}(\phi_1 + \psi_1), 
 \; \theta^2 = {4\,y^2 \over \sqrt{6}L^2}, 
 \; \theta_1^2  =  {8\,x^2 \over \sqrt{6}L^2}, 
 \; \rho^2 = {8\,r^2 \over 3\sqrt{6} L^2}. 
\end{eqnarray} 
Implicitly we have performed a coordinate change from $(\p_1, 
\psi_1,\psi)$ to $(x^+ \propto \p_1+\psi_1, \hat{\psi}, 
\hat{\psi}_1)$ which rotates, as function of the affine parameter, 
the transverse space to the geodesic used to take the Penrose 
limit.  This eliminates any "magnetic" terms in $\hat\psi_1$ and 
$\hat\psi$ in the final form of the metric for the Penrose limit 
in the Pilch--Warner background, which can now be written as: 
 \begin{eqnarray} 
 \label{PPIR2} 
 ds^2_{PP} &=& -4dx^+dx^- -{8\over 3}\mu\, [\vec r^2 
+ 4 \vec y^2 + \vec x^2](dx^+)^2 + (d\vec r)^2 +( d\vec y)^2 + ( 
d\vec x)^2 + \emph{O}\,(L^{-1}) \nonumber \\ 
 {\cal A}_{(2)} &=&B_{NS-NS}+iB_{RR} \nonumber \\ 
 &=& a\;d\bar{z}_2\wedge dz_3 + b\;\bar{z}_2\,dx^+\wedge dz_3 
 + c\;z_3\,dx^+\wedge d\bar{z}_2 
 \end{eqnarray} 
where $a$,$\,b$ and $c$ are constants which we were not quite able 
to extract from ref.\cite{pw}.  The momenta  for $x^+$ and $x^-$ 
are: 
 \begin{eqnarray} 
 p_+ &=& \partial_{x^+} 
     = {4 \over \sqrt{6}} \mu ( \partial_{\tau} 
     + {1\over 2} \,\partial_{\phi_1} 
     - {1\over 2} \,\partial_{\psi} 
     + \partial_{\psi_1}) 
 \\ 
 p_- &=& \partial_{x^-} 
     = \;\;{2 \over 3\mu}\; ( \partial_{\tau} 
     - {1\over 2} \,\partial_{\phi} 
     + {1\over 2} \,\partial_{\psi} 
     - \partial_{\psi_1}) 
 \end{eqnarray} 
 
 Note that in both the solutions above we normalize $x^+$ such that 
 when these pp-wave solutions appear as IR endpoints for the 
 full RG-flow pp-wave, $x^+$ is continued into the UV region 
 such that 
  \be 
   {\partial{\psi}\over \partial{x^+}} = -\mu 
  \ee 
 for the $\theta = {\pi\over 2}$ flow and 
  \be 
   {\partial{(\phi_1\pm\psi_1})\over \partial{x^+}} = \mu. 
  \ee 
 for the $\theta = 0$ flows. 
 
   For example, in the $\theta = {\pi\over 2}$ flow 
 there exists a constant of motion $p_{\psi}$ which takes the form 
 \begin{eqnarray} 
 p_{\psi} &=& g_{\psi\psi}\,{\partial{\psi}\over \partial{x^+}} 
 \nonumber\\ 
          &=& {L^2\, \textrm{cosh}\,\chi(r)\over \rho(r)^6} 
          \,{\partial{\psi}\over \partial{x^+}} 
 \end{eqnarray} 
 (see appendix and note that $L^2$ has been absorbed in $\mu$) so 
 \begin{eqnarray} 
   {\partial{\psi}\over \partial{x^+}}|_{UV} &=& {1\over 
   L^2}\,p_{\psi}= -\mu \nonumber \\ 
   {\partial{\psi}\over \partial{x^+}}|_{IR} &=& {\sqrt{3}\over 
   L^2}\,p_{\psi}= -\sqrt{3}\,\mu 
 \end{eqnarray} 
 which normalizes eq. (\ref{var1}) just right. 
 
\begin{figure}[th] 
 \centerline{\psfig{figure=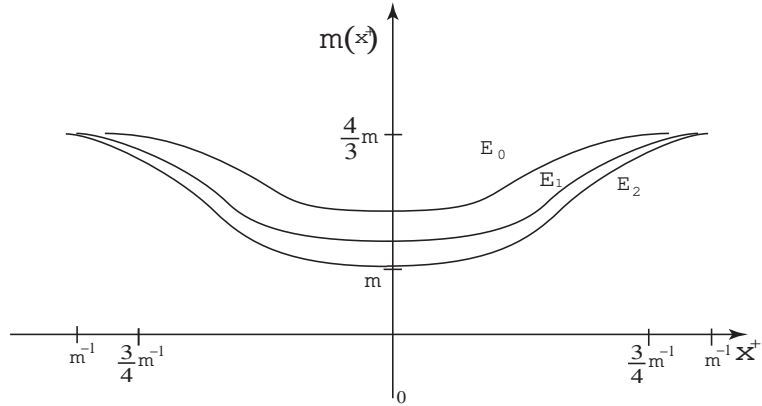,width=10cm,clip=}} 
 \caption{$\mu(x^+)$ for various values of $E$; $E_0 < E_1 < E_2$} 
 \label{fig:mufig} 
\end{figure}

Whereas for the two fixed points we have a ``global'' coordinates, 
at any generic point along the flow we have a formulation only in 
terms of the Poincar\'{e} coordinates. In the appendix we briefly 
summarize the determination of the masses for the Penrose limit 
taken in Poincar\'{e} coordinates. Following that discussion we 
know that the mass--squared associated with the $AdS_5$ 
coordinates are given by 
 \be -\mu_x^2 = {\partial^2_{x^+} \sqrt{g_{tt}}\over 
\sqrt{g_{tt}}}={\partial^2_{x^+} \Omega e^A\over \Omega e^A} = 
1/2\pa_r(\dot r)^2[{\pa_r \Omega\over \Omega}+ \pa_r A] + (\dot 
r)^2[{\pa^2_r \Omega\over \Omega}+ \pa^2_r A + 2{\pa_r 
\Omega\over \Omega}\pa_r A + (\pa_r A)^2]] 
 \ee 
Since there is no analytic expression for $\chi(r)$ and $\tilde 
\rho(r)$ we can write down an analytic expression for the masses 
along the whole flow. However, it is easy to check that $\mu_x^2$ 
is $\mu^2$ and $4/3 \mu^2$ for the UV and IR fixed points 
respectively, where we have used the geodesic line associated 
with (\ref{PWp2}).  The function $\mu_x^2(x^+)$ for the 
$\theta={\pi\over 2}$ limit will have the functional form in 
fig.\ref{fig:mufig} for various values of E (the conjugate 
momentum to $t$). 
 
%%%%%%%%%%%%%%%%%%%%%%%%%%%%%%%%%%%%%%%%%%%%%%%%%%%% 
\subsection{Flows in the corresponding field theory} 
%%%%%%%%%%%%%%%%%%%%%%%%%%%%%%%%%%%%%%%%%%%%%%%%%%%% 
 
    Now that we have some examples of workable Penrose limits for 
the 10-dimensional PW background describing the flow from $N=4$ 
SYM to the $N=1$ SCFT with 2 adjoints, we will examine what these 
limits mean within the context of the field theory.  The first 
step in establishing this understanding is to set down how the 
symmetries of the "sphere" part  of the SUGRA solution map to 
R-symmetries in the field theory. 
 
    The original R-symmetry of $N=4$ SYM is $SU(4)$.  The 3 
complex adjoint fields $\Phi_1,\Phi_2,\Phi_3$ transform in the 
{\bf 6} representation of this group, while the spinors 
$\Psi_1,\Psi_2,\Psi_3,\Psi_4$ transform in the {\bf 4}. We start 
the flow by introducing a mass term for $\Phi_3$ and $\Psi_3$, and 
now the flow only preserves an $SU(2)\times U(1)_R$ subgroup of 
$SU(4)$ with the following charge assignment: 
 
%\begin{table*}[h] 
\begin{center} 
\begin{tabular}[h]{|c|c|c|} 
\hline 
                   & $T_3$ & $U(1)_R$  \\ 
 
\hline 
$(\Phi_1,\Psi_1)$ & 1/2  & $(1/2,-1/2)$ \\ 
\hline 
$(\Phi_2,\Psi_2)$ & -1/2 & $(1/2,-1/2)$ \\ 
\hline 
$(\Phi_3,\Psi_3)$ &   0  & $(1,0)$  \\ 
\hline 
$\Psi_4$          &   0  & $1$  \\ 
\hline\end{tabular} 
\end{center} 
%\end{table*} 
with $T_3$ generating the $U(1)$ of $SU(2)$. 
 
    In the metric (\ref{bigmetric}) of the PW SUGRA solution, we see that there exist 
an $SU(2)_L\times U(1)_{\phi_1}\times U(1)_\psi$ symmetry: the 
forms $\sigma^i$ are by definition invariant under $SU(2)_L$ and 
the metric is explicitly independent of the angles $\phi_1$ and 
$\psi$. If we align our definition of the $\Phi_i$'s with the 
$u_i$'s used in the derivation of the PW metric, we are lead to 
the following charges: 
 
%\begin{table*}[h] 
\begin{center} 
\begin{tabular}[h]{|c|c|c|c|} 
\hline 
                  & $T_3$ & $U(1)_{\phi_1}$ & $U(1)_{\psi}$ \\ 
\hline 
$(\Phi_1,\Psi_1)$ & 1/2  & $(1/2,0)$  & $(0,1/2)$   \\ 
\hline 
$(\Phi_2,\Psi_2)$ & -1/2 & $(1/2,0)$  & $(0,1/2)$   \\ 
\hline 
$(\Phi_3,\Psi_3)$ &   0  & $(0,-1/2)$ & $(-1,-1/2)$ \\ 
\hline 
$\Psi_4$          &   0  & $+1/2$     &  $-1/2$     \\ 
\hline\end{tabular} 
\end{center} 
%\end{table*} 
from which we conclude that 
 \be 
 \label{U1R} 
 U(1)_R = U(1)_{\phi_1} - U(1)_{\psi} = -i\,(\partial_{\phi_1} + 
 \partial_{\psi}). 
 \ee 
  A quick look at the form of $A_{(2)}$ in the appendix (\ref{A2}) 
confirms this picture, as the complex 2-form transforms under the 
combination $(\phi_1 + \psi)$, which leaves $U(1)_R$ alone and 
breaks the other combination of the two $U(1)$'s. Having now 
identified the various charges of fields, we can take a look at 
what flows lead to the two PP-wave limits above. 
 
\subsubsection{The $\theta = {\pi \over 2}$ Flow} 
 
    In the IR limit, we took the $\theta = {\pi \over 2}$ Penrose 
limit with 
 \be 
 \label{IR1} 
 x^+ = {\sqrt{3} \over 4\mu}(\tau - {2\over 3}\,\psi) 
 \qquad\textrm{and}\qquad 
 p_+ = {2\over \sqrt{3}}\,\mu\,( \partial_{\tau} 
     - {3\over 2} \,\partial_{\psi} 
     - {1\over 2} \,\partial_{\psi_1}). 
 \ee 
When this IR fixed point is embedded in the PW flow solution the 
natural choice of variables which continuously connects the affine 
parameter (see appendix) to the UV gives in that limit 
 \be 
 \label{UV1} 
 x^+ = {1\over 2\mu}(\tilde{\tau} - \psi) \qquad\textrm{and}\qquad 
 p_+ = \mu\,(\partial_{\tilde{\tau}} - \partial{\psi}). 
 \ee 
The variables $\tau$ and $\tilde{\tau}$ are only well defined in 
the IR and UV respectively, where we recover conformal invariance. 
At the two fixed points, conformal invariance implies that the 
bulk space approaches $AdS_5$ for which we can define ``global 
time'' with conformal killing vectors $\partial_{\tau}$ and 
$\partial_{\tilde{\tau}}$ respectively.  It is still unclear to us 
whether or not one can define a good coordinate on the whole flow 
which reduces to ``global time'' at each of the fixed points, but 
such a coordinate will not be necessary for our purposes. 
 
We now summarize in the following tables the properties of the 
chiral superfields in the IR and UV ends. For the ${\cal N} = 4$ 
UV theory, eq. (\ref{UV1}) tells us that $J = -U(1)_{\psi}$.  This 
is the $U(1)$ subgroup of $SU(4)_R$ which acts naturally on 
$\Phi_3$: 
 
%\begin{table*}[h] 
\begin{center} 
\begin{tabular}[h]{|l||c|c|c|c|} 
\hline 
       {\bf UV}                 & $SU(N)$   &     $J$       & $\Delta$     & $\Delta-J$\\ 
\hline \hline 
$(\Phi_3,\Phi_3^{\dag},\Psi_3)$ & $adjoint$ &  $(1,-1,1/2)$ & $(1,1,3/2) $ & $(0,2,1)$ \\ 
\hline \hline 
$(\Phi_1,\Phi_1^{\dag},\Psi_1)$ & $adjoint$ &  $(0,0,-1/2)$ & $(1,1,3/2) $ & $(1,1,2)$ \\ 
\hline 
$(\Phi_2,\Phi_2^{\dag},\Psi_2)$ & $adjoint$ &  $(0,0,-1/2)$ & $(1,1,3/2) $ & $(1,1,2)$ \\ 
\hline 
$(D_i\Phi_3,D_i\Psi_3)$         & $adjoint$ &  $(1,1/2)$    & $(2,5/2) $   &  $(1, 2)$ \\ 
\hline\end{tabular} 
\end{center} 
%\end{table*} 
 
For the ${\cal N} = 1$ IR theory, we can read from eq. 
(\ref{IR1}) that $J = -{3\over 2}U(1)_{\psi} - {1\over 2} 
U(1)_{\phi_1}$.  This is not a conserved current of the ${\cal 
N}=1$ SCFT at the quantum level, which implies that the pp-wave 
limit must restore some extra supersymmetry. 
 
%\begin{table*}[h] 
\begin{center} 
\begin{tabular}[h]{|l||c|c|c|c|} 
\hline 
       {\bf IR}                 &$SU(N)$  &     $J$         & $\Delta$        & $\Delta-J$  \\ 
\hline \hline 
$(\Phi_3,\Phi_3^{\dag},\Psi_3)$ &$adjoint$& $(3/2,-3/2,1)$  & $(3/2,3/2,2)$   & $(0,3,1)  $ \\ 
\hline \hline 
$(\Phi_1,\Phi_1^{\dag},\Psi_1)$ &$adjoint$&$(-1/4,1/4,-1/2)$& $(3/4,3/4,5/4)$ & $(1,1/2,2)$ \\ 
\hline 
$(\Phi_2,\Phi_2^{\dag},\Psi_2)$ &$adjoint$&$(-1/4,1/4,-1/2)$& $(3/4,3/4,5/4)$ & $(1,1/2,2)$ \\ 
\hline 
$(D_i\Phi_3,D_i\Psi_3)$         &$adjoint$&    $(3/2,1)$    & $(5/2,3) $      & $(1, 2)$ \\ 
\hline\end{tabular} 
\end{center} 
%\end{table*} 
 
    From these two tables it seems clear that the proper operator 
to expand around is Tr$[\Phi_3^J]$.  It will provide a natural 
vacuum state for the dual string theory with 
$\Phi_1,\Phi_1^{\dag},\Phi_2,\Phi_2^{\dag}$ and $(D_i\Phi_3)$ 
acting as the oscillator insertion operators.  It is important 
note that in the IR, the operators $\Phi_1^{\dag}$ and 
$\Phi_2^{\dag}$ have $\Delta - J = 1/2$ as opposed to $1$ as 
might be expected from the form of the metric (\ref{pimetric}). 
The difference can be explained on the supergravity side by the 
presence of a non-zero time-dependent $B_{\mu\nu}$ field which 
will modify the energy of some of the string states.  Since such 
a $B$--field only couples to the world sheet with a $\sigma$ 
derivative, it is not yet clear to us how the level--0 oscillator 
energies are shifted.

\subsubsection{The $\theta = 0$ Flow} 
 
In the $\theta=0$ limit we get in the IR an affine parameter with 
the property that 
 \be 
 \label{IR2} 
 x^+ = {\sqrt{6} \over 8\mu}(\tau + {2\over 3}\,(\phi_1+\psi_1)) 
 \qquad\textrm{and}\qquad 
 p_+ = {4\over \sqrt{6}} \,\mu\,( \partial_{\tau} 
     + {1\over 2} \,\partial_{\phi_1} 
     - {1\over 2} \,\partial_{\psi} 
     + \partial_{\psi_1}). 
 \ee 
If we continue this choice into the UV along a geodesic in the 
PW-flow with large energy ($\partial_t$) we get in the UV: 
 \be 
 \label{UV2} 
 x^+ = {1\over 2\mu}(\tilde{\tau} - \psi) \qquad\textrm{and}\qquad 
 p_+ = \mu (\partial_{\tilde{\tau}} + \partial_{\phi_1} + \partial{\psi_1}). 
 \ee 
For this choice of $J$ in the ${\cal N}=4$ UV limit corresponds 
to the $U(1)$ which acts naturally on $\Phi_1$.  We can then draw 
a table of charges in the UV very similar to that related to the 
$\theta = {\pi \over 2}$ geodesics: 
%\begin{table*}[h] 
\begin{center} 
\begin{tabular}[h]{|l||c|c|c|c|} 
\hline 
       {\bf UV}                 & $SU(N)$   &     $J$       & $\Delta$     & $\Delta-J$\\ 
\hline \hline 
$(\Phi_1,\Phi_1^{\dag},\Psi_1)$ & $adjoint$ &  $(1,-1,1/2)$ & $(1,1,3/2) $ & $(0,2,1)$ \\ 
\hline \hline 
$(\Phi_2,\Phi_2^{\dag},\Psi_2)$ & $adjoint$ &  $(0,0,-1/2)$ & $(1,1,3/2) $ & $(1,1,2)$ \\ 
\hline 
$(\Phi_3,\Phi_3^{\dag},\Psi_3)$ & $adjoint$ &  $(0,0,-1/2)$ & $(1,1,3/2) $ & $(1,1,2)$ \\ 
\hline 
$(D_i\Phi_1,D_i\Psi_1)$         & $adjoint$ &  $(1,1/2)$    & $(2,5/2) $   &  $(1, 2)$ \\ 
\hline\end{tabular} 
\end{center} 
%\end{table*} 
We can use eqs. (\ref{IR2}) and (\ref{U1R}) to derive that the 
appropriate $J$ in the IR is 
 \be 
 J = {1\over 2} U(1)_R + T_3. 
 \ee 
which allows us to read off the relevant charges in the IR: 
 
\begin{center} 
\begin{tabular}[h]{|l||c|c|c|c|c|c|} 
\hline 
                   & $SU(N)$  & $T_3$ & $U_R(1)$ & $\Delta$& $\Delta-J$  \\ 
\hline \hline 
$(\Phi_1,\Phi_1^{\dag},\Psi_1)$&$adjoint$&$(1/2,-1/2,1/2)$&$(1/2,,-1/2,-1/2)$&$(3/4,3/4,5/4)$&$(0,2,1)$ \\ 
\hline \hline 
$(\Phi_2,\Phi_2^{\dag},\Psi_2)$&$adjoint$&$(-1/2,1/2,-1/2)$&$(1/2,-1/2,-1/2)$ & $(3/4,3/4,5/4) $ &  $(1,1/2,2)$ \\ 
\hline 
$(\Phi_3,\Phi_3^{\dag},\Psi_3)$ & $adjoint$ & 0    &  $(1,-1,0)$      & $(3/2,3/2,2) $ &  $(1,2,2)$ \\ 
\hline 
$(D_i\Phi_1,D_i\Psi_1)$ & $adjoint$ & 1/2 &  $(1/2,-1/2)$ & $(7/4,9/4) $ & $(1, 2)$ \\ 
\hline\end{tabular} 
\end{center} 
 
Note that we now have Tr$[\Phi_1^J]$ as the operator dual to the 
string vacuum state throughout the whole flow.  The oscillators 
will then correspond to 
$\Phi_2,\Phi_2^{\dag},\Phi_3,\Phi_3^{\dag}$ and $(D_i\Phi_1)$. 
Since the pp-wave metric in eq. (\ref{PPIR2}) has twice the mass 
for the direction along $\vec{y}$ we expect that the two 
oscillators which correspond to this direction will have 
$\Delta-J=2$ in the IR. These oscillators have 
$-\partial_{\hat{\psi}} = \pm 1$ and so can be readily identified 
as $\Phi_3$ and $\Phi_3^{\dag}$.  Looking at the table above we 
see that relative to the metric there are two fields with 
``deviant'' $\Delta-J$: $\Phi_3$ and $\Phi_2^{\dag}$.  This 
difference should be accounted for after the $B_{\mu\nu}$ field 
effects on the worldsheet are accounted for. 
 
\subsubsection{General Comments on the limits}

    For both of the flows above, if we consider $p_+$ at a generic 
point in the flow, we will not be able to expand in terms of the 
vector field $\partial_{\tau}$ since it is  only defined at the 
fixed points.  However, we can always isolate the $S^5$ part of 
the vector $p_+$.  It is possible to see what this might be 
proportional to by looking carefully at how $J$ is defined in the 
UV and IR for both the flows.  We get that 
 \begin{eqnarray} 
 J_{IR} &=& J_{UV} - {1\over 2}(U(1)_{\phi} + U(1)_{\phi1}) \nonumber \\ 
        &=& J_{UV} - \gamma^i_{IR}\;K^i 
 \end{eqnarray} 
where the $\gamma^i$ are the corrections to the anomalous 
dimensions for the $i^{th}$ chiral superfield, and $K^i$ is the 
Konishi current which assigns charge 1 to the $i^{th}$ chiral 
superfield. This means that a natural guess that the $S^5$ related 
part of $p_+$ is proportional to $J_{UV} - \gamma^i(x^+)\;K^i$.

As a last remark of this section let us following the discussion 
in section 2 compute the transition amplitude between the initial 
and final ground states. Substituting the values of $\mu$ for 
instance for the $\theta=\pi /2$ case, $\mu^{(i)}=1, 
\mu^{(f)}=2/\sqrt{3}$ into  (\ref{sudden}) we find that the 
transition probability is .997. 
 
% derivation for the ${\cal N}=1$ end-point the 
%value of the light-cone time takes the form $x^+=\psi+\phi_1$ 
%which translates into $P^- \alpha \Delta -J$ where $J= \half R + 
%T_3$, where we used the fact that $U(1)_R$ is conjugate to 
%$\phi_1 - \psi$ . The flow from the ${\cal N}=4$ end-point to 
%that of the 
% ${\cal N}=1$  can be along a trajectory of $P^-$ just defined, namely 
% $P^-(\Lambda)= \Delta(\Lambda) -J(\Lambda)$ where $\Lambda$ is the RG scale which 
%translate in the world sheet Hamiltonian discussed in (..) to 
%$H_{ws}(\tau)= \Delta(\tau) -J(\tau)$. Following this picture we 
%would like to have a Penrose limit of the ${\cal N}=4$ which 
%still associate with the same $J= \half R + T_3$. 
 
%Recall that in the  ${\cal N}=4$ case the $U(1)_R$ is defined via 
%the decomposition of $SU(4)\rightarrow SU(3)\times U(1)_R$. In 
%that case the operators of above take the following values in the 
%${\cal N}=4$  end-point. 
 
%\subsection{ The supersymmetry algebra} 
%\end{itemize} 
 
\section{The pp--wave limit of large N $D_p$ branes} 
 
%%%%%%%%%%%%%%%%%%%%%%%%%%%%%%%%%%%%%%%%%%%%%%%%%%%%%%%%%%%%%% 
 
%%%%%%%%%%%%%%%%%%%%%%%%%%%%%%%%%%%%%%%%%%%%%%%%% 
\subsection{Dp branes and their Penrose Limit} 
%%%%%%%%%%%%%%%%%%%%%%%%%%%%%%%%%%%%%%%%%%%%%%%% 
 
Before proceeding to study the \pg\ of $D_p$ we briefly 
review the corresponding backgrounds. In  the field 
theory limit of $D_p$ branes, which was  discussed in 
\cite{cobi} we have 
 
\be\label{limit}  \gym^2=(2\pi)^{p-2} 
g_s\a^{(p-3)/2}=\mbox{fixed},\;\;~~~~~  \a \to  0, 
\ee 
where 
$g_s=e^{\phi_{\infty}}$, and $\gym $ is the Yang-Mills  coupling 
constant. The energies are kept fixed when we take the  limit. For 
$p\leq3$ this limit implies that the theory decouples from  the bulk 
since the ten dimensional Newton constant goes to zero.  It also 
suppresses  higher order corrections in $\alpha'$ to the  action. At a 
given energy scale, $U$, (related to the radial coordinate  as 
$U\equiv \frac{r}{\a} $) the effective  dimensionless  coupling 
constant in the corresponding super-Yang-Mills theory  is 
$g^2_{eff}\approx \gym^2NU^{p-3}$. 
This implies that  weakly coupling 
calculations in SYM are valid if 
\be 
g^2_{eff} \ll 1 
~~~~\Rightarrow~~~~ \left\{ \ba{l}  U\gg (\gym^2 N)^{1/(3-p)}~,~~~~p<3 
\\    U \ll 1/(\gym^2 N)^{1/(p-3)}~,~~~~p>3 
\ea 
\right. 
\ee 
In the 
Maldacena limit the supergravity solution describing a  stack of $N$ 
$D_p$ branes is 
\bea 
\label{gsol} 
&& ds^2=\a\left( 
\frac{U^{(7-p)/2}}{\gym\sqrt{c_p N}}dx^2_{||}+ \frac{\gym\sqrt{c_p 
N}}{U^{(7-p)/2}}dU^2+\gym\sqrt{c_p  N}U^{(p-3)/2} 
d\Omega^2_{8-p}\right) , \non 
&&  e^{\phi}=(2\pi)^{2-p} \gym^2\left( 
\frac{\gym^2 c_p  N}{U^{7-p}}\right) ^{\frac{3-p}{4}} \sim 
\frac{\gef^{(7-p)/2}}{N}. 
\eea 
where $c_p = 2^{7-2p} \pi^{9-3p \over 
2}  \Gamma({7-p \over 2}) $.  Note that  the effective string 
coupling, $e^{\phi}$, is finite  in the decoupling limit. In terms of 
$\gef$ the curvature  associated with the metric (\ref{gsol}) is 
\be 
\label{curg} 
\a R\approx {1\over \gef } \sim \sqrt{{ U^{3-p} \over 
\gym^2 N }}. 
\ee 
Since from the field theory point of view $U$ is an 
energy scale,  going to the UV in the field theory means taking the 
limit  $U \to \infty$. In this limit we see from eq.(\ref{gsol})  that 
for  $p<3$ the effective string coupling vanishes and the theory 
becomes UV free. For $p>3$ the coupling increases and we  have to go 
to a dual description  before we can investigate it reliably. This 
property  of the supergravity solution is closely related to  the fact 
that for $p>3$ the  super-Yang-Mills theories are  non-renormalizable 
and hence, at short distances, new degrees of  freedom appear.

In order to trust the type II supergravity solution (\ref{gsol})  we 
need both the curvature (\ref{curg})  and  the dilaton  (\ref{gsol}) 
to be small. This implies $  1 \ll  \gef^2 \ll N^{4 \over 7-p}$.

Since the main intuition in the  understanding of Penrose limits 
arises from the study of the  near horizon limit of a stack of $D3$ 
branes, we choose to work in a set of coordinates that shares some 
properties with  $AdS_5\times S^5$ \cite{skenderis,magoo} and is valid 
for $p<5$ 
 
\bea 
\label{z} 
ds^2 &=&  l_s^{7-p\over 5-p}R^{4\over 
  5-p}z^{3-p\over 5-p}\bigg[{-dt^2+d\vec{x}^2+ dz^2 \over 
 z^2}+{(5-p)^2\over 4}d\O_{8-p}^2\bigg], \non 
e^{\phi}&\sim&{1\over N}l_s^{-(7-p)(3-p)\over 2(5-p)} 
 R^{2(7-p)\over (5-p)} 
 z^{(7-p)(3-p)\over 2(5-p)}, \non 
A_{0...p} &=&  \half\left [ 1-\left( R^{8\over 5-p} z^{2(p-7)\over p-5}\right)^{-1}\right ]. 
\eea 
The $z$ 
coordinate used here is related to the more conventional $U$ coordinate 
according to 
 \be 
z={2\sqrt{c_p \gym^2 N}\over (5-p)U^{5-p\over 2}}. 
 \ee 
 
In order to understand the relevance of the Penrose limit for the 
dual gauge theory it was important in the BMN construction to 
perform the limit by taking the radius of AdS to infinity; in the 
field theory it implied taking the large $N$ limit since for 
$AdS_5\times S^5$ we have that the dimensionless number $R^4=4\pi 
g_s N$. It is thus, important for us to elucidate this relation 
in the context of Dp-branes. The dimensionless ``radius'' $R$ is defined by 
 \be 
 R^4= ({2\over 5-p})^{7-p} c_p g_s N. 
 \ee so that it converges to the dimensionless radius of the $AdS_5\times S^5$.
If we naively take 
the $R\rightarrow \infty$ as part of the \pg\   the effective 
string coupling $e^\phi$ diverges for the all the $D_p$ 
backgrounds with $p\leq 5$ (apart from $p=3$). There are 
different approaches to avoid this situation. The simplest one is 
to rescale the  string scale $l_s$ 
 \be 
l_s = \tilde l_s R^{{4\over 3-p}} 
 \ee 
 with finite $\tilde l_s$ so that $l_s$ vanishes for $p<3$ and 
diverges for $p>3$. 
The background (for $p\neq 3$) expressed in terms of $\tilde l_s$ 
now reads 
 \bea\label{metricz} 
ds^2 &=& \tilde  l_s^{7-p\over 5-p}R ^{8\over (3-p) } 
 z^{3-p\over 5-p}\bigg[{-dt^2+d\vec{x}^2+ dz^2 \over 
z^2}+{(5-p)^2\over 4}d\O_{8-p}^2\bigg].\non 
 e^{\phi}&\sim&{1\over N}\tilde l_s^{-(7-p)(3-p)\over 2(5-p)} 
z^{(7-p)(3-p)\over 2(5-p)} \non A_{0...p} &=&  \half\left [ 
1-\left( R^{8\over 5-p} z^{2(p-7)\over p-5}\right)^{-1}\right ] 
 \eea 
From the form of the rescaled metric it is thus clear that the 
\pg\ with $g_sN\rightarrow \infty$ can be taken for only for 
backgrounds of $D_{p\leq 3}$.\footnote{This point of view is 
rather limited but allows us to proceed. A more general approach, 
uniformly treating all $D_p$ branes, will be outlined at the end 
of this subsection since it requires knowledge of properties of 
the geodesic that we are discussing. We thank F. Larsen for 
suggesting this approach to us.} With the intention of applying 
the \pg\  we consider a null geodesic in the plane expanded by 
$(t,z,\psi)$ where $\psi$ an angle in the $(8-p)$--dimensional 
sphere with metric: $ d\psi^2 + \sin^2\psi\ d\O_{7-p}^2$. The 
effective Lagrangian characterizing the motion along that 
geodesic is 
 \be 
\label{lag} 
\LL =-z^{-{7-p\over 5-p}}\dot t^2 + z^{-{7-p\over 5-p}}\dot z^2+ 
z^{3-p\over 5-p}\dot \psi^2, 
 \ee 
where dot means differentiation with respect to the affine 
parameter $\l$ and we have rescaled $\psi$ by $2/(5-p)$. The 
equations of motion resulting from this Lagrangian together with the 
constraint are: 
 \be 
\label{geoeom} 
\dot\psi =\m \, z^{-{3-p\over 5-p}}, \qquad \dot t =E\,\, 
z^{7-p\over 5-p}, \qquad \dot z =\pm z^{2\over 5-p}\sqrt{[ E^2\, 
z^{2}-\m^2]} 
\ee 
Note that the null geodesic line is well defined only for $z\geq {\m\over E}$. 
Our aim, following \cite{penrose} (see also \cite{tratado}) is to 
find coordinates convenient for the Penrose limit, i.e., 
coordinates where the metric takes the form 
 \be 
ds^2=d\b(d\l+a d\b + b_i dY^i) + C_{ij}dY^i dY^j, 
 \ee 
where $a, b_i$ and $C_{ij}$ can depend on any coordinate. For the 
metric  (\ref{metricz}) we need a transformation of the form: 
$(t,z,\psi) \to (\l,\b,\p)$, such that $g_{\l\l}=g_{\l\p}=0$ and 
$g_{\l\b}=1$. The condition $g_{\l\l}=0$ follows directly from 
the fact that $\l$ is the affine parameter of the null geodesic 
determined by (\ref{lag}) since $g_{\l\l}=\dot z^2 G_{zz} + \dot 
t^2 G_{tt}+\dot \psi^2 G_{\psi\psi}=\LL=0$. 
 
The other two conditions on the metric in the new coordinates 
$(g_{\l\b}=1, g_{\l\p}=0)$, imply a first order system of 
differential equations that can be satisfied by 
 \be 
z=z_\l, \quad t=t_\l-{\b \over E} + \m \p, \qquad \psi=\psi_\l + 
E\p, 
 \ee 
where the subindex $\l$ represents dependence only on $\l$. The 
quantities $r_\l, \, t_\l$ and $\psi_\l$ are completely 
determined from equation (\ref{geoeom}), in the general case we 
have the following implicit solution: 
 \bea 
\label{eenn} 
E\left({\m \over E}\right)^{2\over 5-p}\,\, \l &=& \int^{E\,z/\m} 
{dy\over y^{2\over 5-p}\sqrt{y^2-1}}, \non 
\psi_\l&=&-\arctan{1\over \sqrt{{E^2\over \m^2}z^2-1}}, \non 
t_\l&=& {\m \over E} \,\sqrt{{E^2\over \m^2}\, z^2-1}, 
 \eea 
where $y= {E\over \m}z$.  For $p=0,1,2$ the above integral is 
given by Elliptic integral of first kind, for $p=3$ we obtain a 
trigonometric function and for $p=4$ is simply reduces to a 
square--root (see below). As can be seen from previous equations 
(\ref{z})-(\ref{eenn}) we are, in fact, considering a family of 
geodesic characterized by $(E,\m)$. To uniformly treat all 
Dp-branes we need to follow a specific member of this family by 
rescaling $E$ and $\m$ differently. This provides and alternative 
to rescaling $R$ and $l_s$. 
 
The metric in the new coordinates takes the expected form 
 
 \bea 
ds^2&=&2d\l\,d\b-{1\over E^2}z^{-{7-p\over 5-p}}d\b^2 + 2{\m 
\over E}z^{-{7-p\over 5-p}}d\b d\p +\big[E^2\, z^{3-p\over 
5-p}-\m^2 \, z^{-{7-p\over 5-p}}\big]d\p^2 \non &+& z^{-{7-p 
\over 5-p}}(dx_1^2+...+dx_p^2) +{(5-p)^2\over 4}z^{3-p\over 5-p} 
\, \sin^2(\psi_\l + E\p)d\O_{7-p}^2. 
 \eea 
After taking the \pg , which involves taking $\O\to 0$ $(R\rt 
\infty)$ while rescaling  the metric as $\O^{-2}$ with the 
following rescaling of the coordinates 
 \be 
\l=\l, \qquad  \b=\O^2\, \b, \qquad Y^i=\O\,y^i, 
\ee 
the metric becomes 
 \be 
ds^2=2d\l\,d\b +\big[E^2\, z^{3-p\over 5-p} -\m^2 \, 
z^{-{7-p\over 5-p}}\big]d\p^2 + z^{-{7-p \over 5-p}}d\vec{x}^2 
+{(5-p)^2\over 4}z^{3-p\over 5-p} \, \sin^2(\psi_\l)ds^2({\bf 
R}^{7-p}). 
 \ee 
One can introduce the Brinkman or harmonic coordinates in which 
case the metric takes the less explicit form 
 \be 
 ds^2=-4dx^+ dx^- - (m_\p^2 \p^2 
 + m_x^2 \sum\limits_{i=1}^p x_i^2 
 + m_y^2\sum\limits_{i=p+1}^7 y_i^2 )(dx^+)^2 
 +d\p^2 + \sum\limits_{i=1}^p dx_i^2 + \sum\limits_{i=p+1}^7 dy_i^2 
 \ee 
The presence of a covariantly constant killing vector allows to 
impose the light-cone gauge, in which case $m_i^2$ become the mass of 
the corresponding field in the worldsheet theory. The masses 
\bea 
m_x^2&=&{7-p\over 4(5-p)^2}z^{-2{3-p\over 
5-p}}\bigg[-(3-p)E^2 z^2+(13-3p)\m^2 \bigg], \non 
m_y^2&=&{7-p\over 4(5-p)^2}z^{-2{3-p\over 
5-p}}\bigg[-(3-p)E^2 z^2+(p+1)\m^2  \bigg], \non 
m_\phi^2&=& m_x^2 \non 
\eea 
  We thus have $p+1$ coordinates with  ``mass'' $m_x$ 
and $7-p$  coordinates with mass $m_y$. The masses as a function 
of $z$ are shown when discussing some of the Dp branes. Several 
comments are in order: 
 \begin{itemize} 
 \item 
 As expected for $p=3$ we reproduce the standard result that 
 $m_x^2=m_y^2=\m^2$, the behavior for $p=4$ is qualitatively different 
 from $p<3$ the functions $m_x$ and $m_y$ have zero at the origin and 
 increase very rapidly. 
 \item 
  A general property of the quantities $m_x$ and $m_y$ is 
 that for $p<3$ they have zeroes located at 
 \be 
 z_{x}=\sqrt{13-3p\over 3-p}{\m\over E}, \qquad 
 z_{y}=\sqrt{p+1\over 3-p}{\m\over E} 
 \ee 
 \item 
 Since, as discussed above, the geodesic line 
makes sense only for $z\geq {\mu\over E}$, 
 apart from a small region 
where $\sqrt{13-3p\over 3-p}{\m\over E}\geq z\geq {\m\over E}$ 
 for $m_x^2$ 
 and  $\sqrt{p+1\over 3-p}{\m\over E}\geq z\geq {\m\over E}$ for $m_y$, 
the masses--squared are negative as can be seen in figures 4--6. 
 \end{itemize} 
One can check that these negative value of $m^2$ are consistent 
with the SUGRA equations of motions.  As discussed above in 
section 2 the Ricci tensor is related to the masses as follows 
 \be 
 R_{++}= (p+1) m_x^2 + (7-p) m_y^2 
 \ee 
Recall that one of the SUGRA equations of motion takes the form 
 \be 
 R_{MN}= -2\nabla_M\nabla_N\phi + {1\over 96}e^{2\phi}[F_{M...}F_{N}^{...} -{1\over 10}G_{MN}F_{...}F^{...}] 
 \ee 
where $_{...}\ ^{...}$ denotes contracted indices. Applying this 
to $R_{++}$ we have for our case 
 \be 
 R_{++}= -2\pa_+\pa_+\phi + {1\over 96}e^{2\phi}[F_{+...}F_{+...}] 
 \ee 
Had the dilaton been constant, this equation would imply that the 
sum $\sum m^2$ must be positive, however this is not the case here 
since we have a running dilaton. In fact it is easy to check that 
for large $z$ the dilaton term dominates the $F^2$ term so that 
one finds 
 \be 
 R_{++}\rightarrow_{z\rightarrow \infty} 
 -2{(7-p)(3-p)\over (5-p)^2}z^{-2{3-p\over 
 5-p}}E^2 z^2 
 \ee 
which is equal to the asymptotic value of $(p+1) 
m_x^2+(7-p)m_y^2$.

%%%%%%%%%%%%%%%%%%%%%%%%%%%%%%%%%%%%%%%%%%%%%%%%%%%%%%%% 
\subsection{Comments on negative negative masses--squared} 
%%%%%%%%%%%%%%%%%%%%%%%%%%%%%%%%%%%%%%%%%%%%%%%%%%%%%%%%% 
The issue of negative masses is a very interesting and intricate 
one and deserves a detailed study on its own. In general, 
pp-waves with negative masses make for a potentially very rich 
laboratory for exploring string theory in an expanding universe. 
Here we will attempt a much more modest heuristic approach and 
argue in favor of negative mass signaling a breakdown in the 
description for the quantum Hilbert space of asymptotic states. 
 
First let us state that what we mean by negative mass originates 
from the coefficient $+ \mu(x^+)^2 X_i^2 (dx^+)^2$ in the generic 
pp-wave metric in Brinkman coordinates. Note, first of all, that 
this term becomes a negative mass--squared for a worldsheet 
coordinate {\it only} after we take the light-cone gauge, that is 
$x^+=p^+\tau$. In other words, there is a gauge choice involved 
in this statement. 
 
Another word of caution is in order: the tachyonic behavior is 
exclusively in the world sheet theory, that is, we are dealing 
with a tachyon in a $1+1$ dimensional field theory. From the 
spacetime point of view it does not make sense to define mass, as 
the system is not invariant under translation of $X^i$'s.  The 
mass--squared term breaks this symmetry. 
\begin{figure}[th] 
 \centerline{\psfig{figure=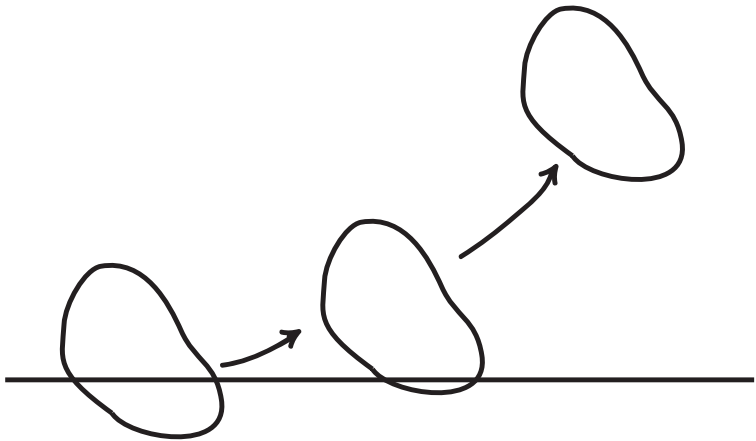,width=10cm,clip=}} 
 \caption{Motion of a stiff string in a background of 
 negative mass squared.} 
 \label{fig:sstring} 
\end{figure} 
 
PP-wave backgrounds with negative mass--squared appear very 
naturally in string theory \cite{horowitz}. In the absence of 
fluxes the Ricci tensor has to vanish and therefore there will be 
masses--squared of opposite signs. Based on the symmetries of 
these backgrounds it was argued in \cite{horowitz} that such 
backgrounds are exact string solutions to all orders in $\a'$. 
Namely, the fact that only $R_{+i+j}$ is nonzero, combined with 
$g^{++}=0$, implies that all curvature tensors necessarily 
vanish. Such scalar tensors are precisely the $\a'$ contribution 
to the SUGRA equations. However, Horowitz and Steif considered 
test strings and found that $<M^2>$ diverges, pointing to a 
quantum mechanical instability. 
 
The clearest analogy (and the main intuition) for the motion of 
the string in these backgrounds is provided by a particle of 
energy $n^2$ moving in a potential given by $\mu(x^+)^2$. The 
geodesic motion along the coordinates $X_i$ with negative mass is 
given by $\ddot X_i-\mu^2 X_i=0$. This means that test masses 
will run away from the origin and feel a tidal force separating 
them of order $\mu^2$. 
 
\begin{figure}[h] 
 \centerline{\psfig{figure=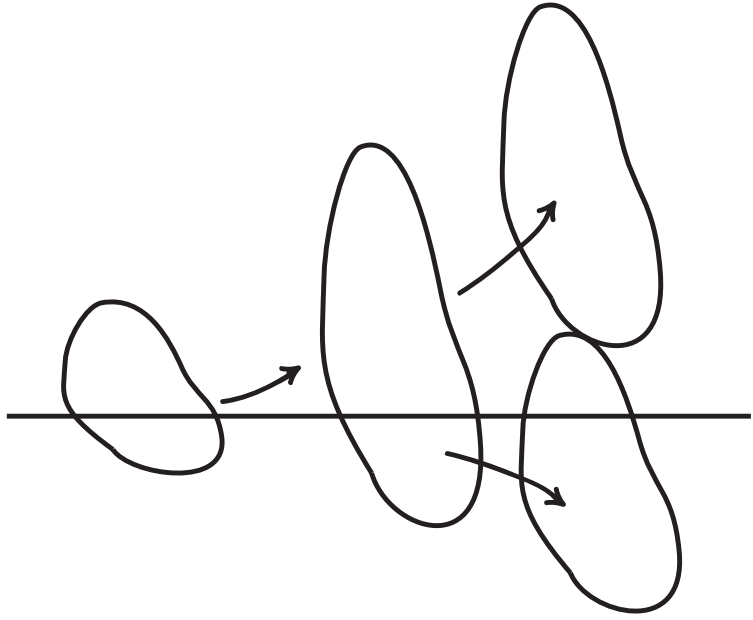,width=10cm,clip=}} 
 \caption{Motion and splitting of a string in a 
 background of large negative mass squared.} 
 \label{fig:lstring} 
\end{figure}

    Coming back to string theory, we see that as long as the string 
tension $\alpha' > \mu^2$, strings will behave like a classical 
particles and nothing unusual happens (see fig.\ref{fig:sstring}). 
Once $\mu^2 > \alpha'$, however, the string will be torn apart due 
to repulsive tidal forces. For any nonzero $g_s$, it will always 
be favorable for the string to split into smaller strings (see 
fig.\ref{fig:lstring}). 
This implies that the Hilbert space of asymptotic strings derived 
from the first-quantized string is no longer valid.  Note that for 
highly excited strings $n^2> \mu^2$ these modes feel nothing. 
 
In the rest of this  section  we describe possible ways to deal 
with the negative mass--squared issue in some of the D--p--brane 
solution. To do so we repeatedly use the D--p--brane maps 
described in \cite{cobi} that point to the most appropriate 
theory to describe the degrees of freedom of a given Dp branes at 
different energy scale.

%%%%%%%%%%%%%%%%%%%%%%%%%%%%%%%%%%%%%%%%%%%%%%%%%%%%%%%% 
\subsection{$p=1$ the flow from $D_1$ to $F_1$} 
%%%%%%%%%%%%%%%%%%%%%%%%%%%%%%%%%%%%%%%%%%%%%%%%%%%%%%%%% 
 
 \begin{figure}[bth] 
 \centerline{\psfig{figure=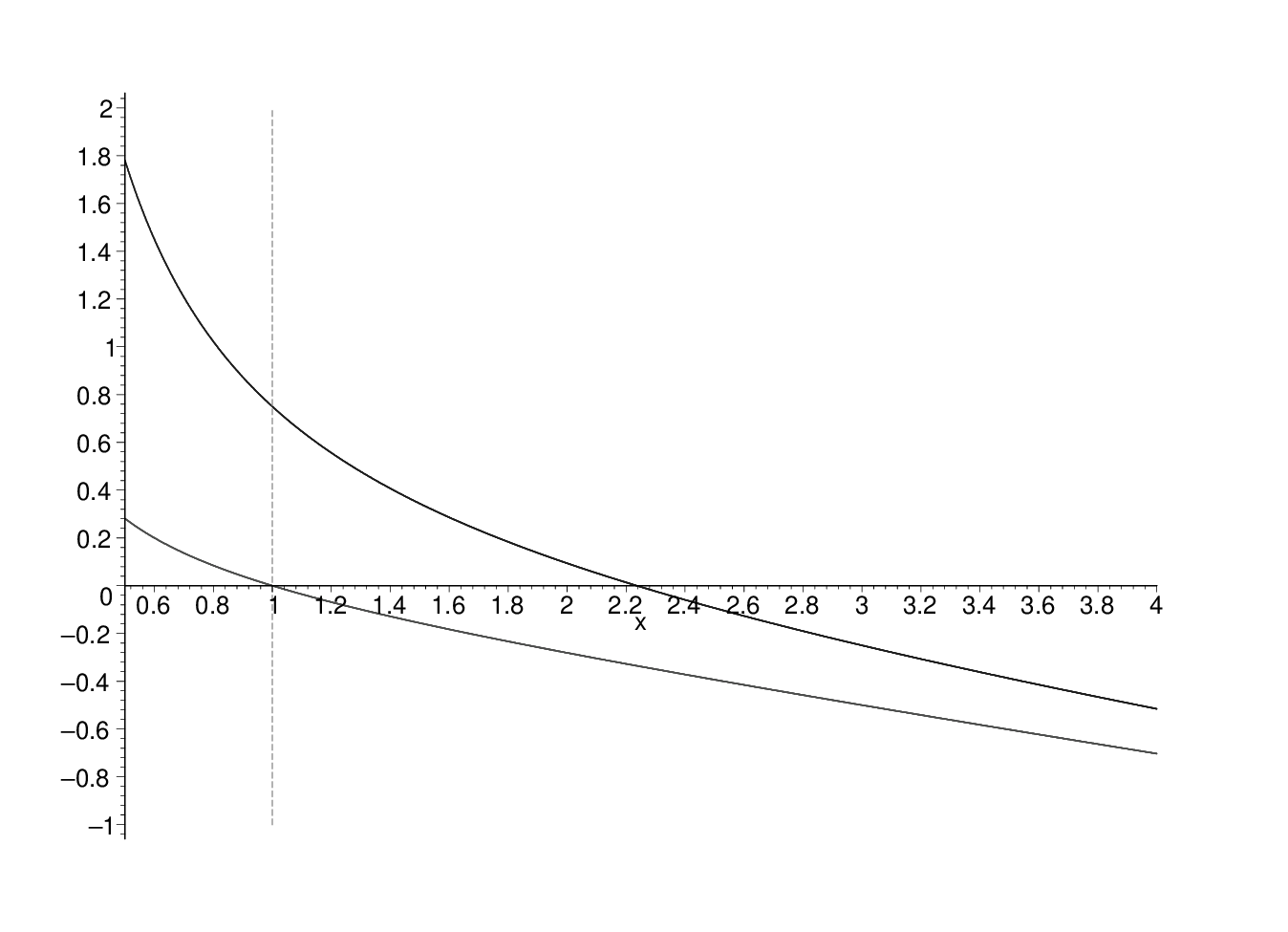,width=11cm,clip=}} 
 \caption{The values of the {\it masses} for D1} 
 \label{fig:d1} 
 \end{figure} 
 
In this case the solution can be written as 
 
\be 
ds^2=-4dx^+ dx^- 
+{3\over 16}z^{-1}\bigg[ 
(E^2z^2-5\m^2)\sum\limits_{i=1}^2 x_i^2 
 + (E^2z^2-\m^2)\sum\limits_{j=3}^8 x_j^2 \bigg](dx^+)^2 
+ \sum\limits_{i=1}^8 dx_i^2, 
 \ee 
where $x^+=\left({2\over E\m}\right)^{1/2}F(\sqrt{{Ez\over 
\m}+1},{1\over \sqrt{2}})$ and $F()$ is a hyper--geometric 
function. As can be seen from figure \ref{fig:d1} the masses of 
this solution are mostly negative and we naturally turn to the 
whole D1-map to understand this situation better. 
 
The $D_1$ SUGRA background \cite{cobi} can be trusted only in the 
${1\over \sqrt {g^2_YM N}} << z <<{1\over \sqrt {g^2_YM N}}N^{2/3}$ otherwise 
either the curvature or the string coupling are not small. For $z$  smaller then the lower bound one has to use the description of 
the $1+1$ SYM theory with 16 supersymmetries. 
In the IR side of the allowed window, where the dilaton becomes large, one 
performs an S-duality transformation and then describe the system  as the near 
horizon limit of large $N$ fundamental strings. 
The Penrose limit of this background can be taken following the same lines as 
for the $D_1$ background. Had we taken the limit on the Einstein metric of the 
fundamental string, obviously the limit would be the same as the one of the $D_1$ string since the Einstein metric does not change under S-duality. 
However, the relevant metric here is obviously the string metric. 
For this metric the geodesic line $\pa_{x^+} z $ and the corresponding mass $m_x^2$ take the form 
 
\bea 
\pa_{x^+} z &=& \pm  {z}^{2}\sqrt{z^2E_{F_1}^2-\mu_{F_1}^2}\\ 
m_x^2&=&{3/4}z^{2}\bigg[E_{F_1}^2 z^2+\m_{F_1}^2 \bigg] 
\eea 
where  the $E_{F_1}$ and  $\m_{F_1}^2$ are the corresponding 
integration constants. 
Recall that for the $D_1$ the geodesic and the mass are given  by 
\bea 
\pa_{x^+} z &=& \pm \sqrt {z}\sqrt{z^2E_{D_1}^2-\mu_{D_1}^2}\\ 
m_x^2&=& 3/16z^{-1}\bigg[-E_{D_1}^2 z^2+5\m_{D_1}^2 \bigg] \eea 
It is thus clear that in the IR region where the $D1$ description 
ceases to be valid, the squared masses become positive again. For 
given parameters $E_{D_1}$ and  $\m_{D_1}^2$ one can guarantee 
continuity in the masses   by adjusting the parameters of the 
string background $E_{F_1}$ and  $\m_{F_1}^2$.

%{\bf to be checked} 
%{\bf add about the transition from the D1 to 
%F1} 
 
 %%%%%%%%%%%%%%%%%%%%%%%%%%%%%%%%%%%%%%%%%%%%%%%%%% 
\subsection{D2 to M2 Flow in 11-d} 
%%%%%%%%%%%%%%%%%%%%%%%%%%%%%%%%%%%%%%%%%%%%%%%%%%% 
 
\begin{figure}[bth] 
 \centerline{\psfig{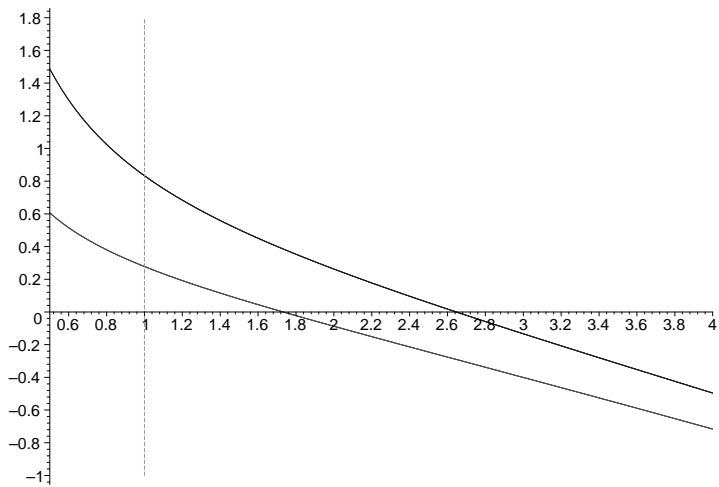}} 
 \caption{The values of the {\it masses} for D2} 
 \label{fig:d2} 
 \end{figure} 
 
As is typical of nonconformal Dp branes a given supergravity 
solution is part of a greater phase diagram called the D2-map in 
\cite{cobi}. Here we review that map since it is crucial for us 
to understand which description is valid at each energy scale. 
Taking the Maldacena limit amounts to 
decoupling the bulk from the theory on the D2 
branes which turns out to be a U(N) super-Yang-Mills in 2+1 
dimensions, with 16 supersymmetries. At a given energy scale, 
$U$, the dimensionless effective coupling  of the gauge theory is 
$g^2_{eff} \sim \gym^2 N/U$ and, hence, perturbative 
super-Yang-Mills  can be trusted in the UV region where $g_{eff}$ 
is small $\gym^2 N \ll U$. 
The type II supergravity description can be trusted when the 
curvature (\ref{curg}) in string units and the effective string 
coupling are small $g^2_{YM} N^{1/5}   \ll  U\ll \gym^2 N$. 
We see that a necessary condition is to have $N \gg 1$. 
In the region $g_{eff}\approx 1$ we have a transition between the 
perturbative super-Yang-Mills description and the supergravity 
description.

In the region $U< \gym^2 N^{1/5}$ the dilaton becomes large. In 
other words the local value of the radius of the eleventh 
dimension, $R_{11}(U)$, becomes larger than the Planck scale 
since $R_{11}=e^{2\phi /3}l_p$. Even though the string theory is 
becoming strongly coupled we will be able to trust the 
supergravity solution if the curvature is small enough in eleven 
dimensional Planck units. 
 
The relation between  the eleven dimensional  metric and the ten 
dimensional type IIA string metric, dilaton and gauge field is 
\be\label{1011} 
ds^2_{11}=e^{4\phi /3 }(dx_{11}+A^{\mu}dx_{\mu})^2+ 
e^{-2\phi /3 }ds^2_{10}, 
\ee 
which implies that the curvature in 11D Planck units is 
\be 
l^2_p R \sim e^{2\phi /3}\frac{1}{g_{eff}}\sim 
\frac{1}{N^{1/3}}\left( \frac{g^2_{YM}}{U}\right) ^{1/3}. 
\ee 
For large $N$, in  the region $\gym^2<U $, the curvature in 11D 
Planck units is small. 
In the region 
  $U<\gym^2$ we should use a different solution 
corresponding to M2-branes  localized on the circle associated 
with the 11th dimension. 
 
 The  curvature of $N$ M2-branes in the field theory limit 
is $l^2_p R \sim \frac{1}{N^{1/3}}$. Note that the 
curvature does not depend on $U$ since the theory is conformal in 
the IR limit. We conclude, therefore, that for large $N$ the 
supergravity description is valid in the region $U\ll \gym^2 
N^{1/5}$.

The M2 brane solution is characterized by a harmonic function $H$ 
and is given by 
 \begin{equation} \label{mtsol} 
 ds^2_{11} = H^{-2/3} dx^2_{||}  + H^{1/3} dx^2_{\perp} 
 \end{equation} 
and there is also a fourform field strength given in terms of $H$. 
%$H$ is a harmonic function. 
When we take $H \sim   N/ x^6$ we have a solution where the M2 
branes are localized in the eight transverse non-compact 
dimensions. If one of the dimensions is compact (let us say the 
11th dimension) we can take $H  \sim N/ r^5$ where now $r$ 
denotes the radial distance in the seven transverse non-compact 
dimensions. This is the solution we get from uplifting 
$N$ D2's.  The appropriate solution to think of interpolating between 
uplifted D2 and M2 is the one in which we take the M2 branes to be 
localized in the compact dimension so that the harmonic function 
is 
 \be\label{superp} 
 H = \sum_{ n = - \infty}^{\infty}  { 2^5 \pi^2 
 N l_p^6 \over ( r^2 + (x_{11} -x_{11}^0 + 2 \pi n 
 R_{11})^2)^3} 
 \ee 
with $x_{11} \sim x_{11} + 2 \pi R_{11} $. For distances much 
larger than $R_{11}$ we can Poisson resum this expression to 
 \be 
 \label{resumed} 
 l_p^3 H = {6 \pi^2 \gym^2 N \over U^5 } + 
 \sum_{m = 1 }^{\infty}  N e^{ - m U/\gym^2 
   }\cos(mx_{11}^0/R_{11}) {\cal O}(U^{-5}), 
 \ee 
where we have used that $R_{11} = g^{2}_{YM} {\alpha'} $. 
%Notice that the exponent is independent of $U$, 
For  $ \gym^2 \ll U$ we can, therefore, use the uplifted solution 
to describe the physics while for smaller values of $U$ we should 
use (\ref{superp}). Note that for such small energy scales it 
becomes necessary to specify the expectation value of $\phi^{11} 
= \gym^2 x_{11}^0/R_{11} $ which is the new scalar coming from 
dualizing the vector in 2+1 dimensions.  For very low  energies 
 \be\label{ir} 
 U \ll \gym^2, 
 \ee 
we are very close to the M2 branes and we can neglect the 
``images'' in (\ref{superp}). Thus,  the solution will resemble 
that of M2 branes in non-compact space and we have the conformal 
field theory with SO(8) symmetry. 
 
The effective Lagrangian describing the motion along null geodesics 
for a metric of the form (\ref{mtsol}) is 
 
 \be 
 {\cal L}=-H^{-2/3}\dot{t}^2+H^{1/3}\dot{r}^2 + H^{1/3}r^2\dot{\psi}^2. 
 \ee 
The corresponding equations of motion 
 \be 
 \dot{\psi}=\mu\,H^{-1/3}, \qquad \dot{t}=E\,H^{2/3}, \quad 
 \dot{r}^2+{\mu^2\over H^{2/3}r^2}=E^2H^{1/3}. 
 \ee 
The coordinate transformation that allows to perform the \pg\ 
limit is, 
 \be 
 r=r(u), \qquad t=t_u-{v\over E}+\m\, \phi, \qquad \psi=\psi_u +E\, \phi, 
 \ee 
where the subindex $u$ means that the corresponding term is a function of 
only $u$. 
After the Penrose limit the metric in Rosen coordinates takes the following 
form 
 \be 
 ds^2=2du dv +(E^2\, H^{1/3}-\mu^2 H^{-2/3})d\phi^2 + H^{-2/3}(dx_1^2 
 +dx_2^2)+ H^{1/3}\big[ r^2 \sin^2\psi_u ds^2({\bf R}^5)+R_{11}^2 dz^2\big]. 
 \ee 
In the more familiar Brinkman coordinates 
 \bea 
 ds^2&=&-4dx^+dx^- - \big[\mu_\phi^2 \phi^2 + \mu^2_p( x_1^2 
 +x_2^2 )+ 
 \mu^2_y(\sum\limits_{i=1}^{5}y_i^2) + \mu^2_z z^2\big](dx^+)^2 \non 
 &+& d\phi^2 + 
 \sum\limits_{i=1}^2dx_i^2 + \sum\limits_{i=1}^5 dy_i^2 +dz^2, \non 
 \mu^2_p&=& {\mu^2\over 3\, r}\sqrt{{E^2r^2H\over \mu^2}-1} 
 {d\over dr}\left({H'\over H^{5/3}\,r}\sqrt{{E^2r^2H\over \mu^2}-1}\right), \non 
 \mu_y^2&=&{\mu^2 \over H^{2/3}r^4}-{\mu^2\over H^{1/2}\, r^2} 
 \sqrt{{E^2r^2H\over \mu^2}-1} 
 {d\over dr}\left({(H^{1/6}r)'\over H^{1/3}\,r} 
 \sqrt{{E^2r^2H\over \mu^2}-1}\right). 
 \eea 
The above solution is not very illuminating. For our purpose the most 
important information is contained in the limits. 
In the M2 limit $(N/r^6)$ we naturally recover the expected 
result $\mu_p^2 =4\mu^2, \,\, \mu_y^2 =\mu^2$. In the uplifted D2 
limit 
 \bea 
 \mu^2_p&=& {20 \mu^2\over L^{10/3} \, r^{2/3}}+ 
 {5\,E^2\, L^{5/3} \over 18\, r^{11/3}}, \non 
 \mu^2_y&=& {35 \mu^2\over 36 \,L^{10/3} \, r^{2/3}}+ 
 {E^2\, L^{5/3} \over 9\, r^{11/3}}, 
 \eea 
The key observation we want to make is that the masses are 
nowhere negative.This result shows that once lifted 
to M-theory the possible instability is gone.

%{\bf add about the transition from negative masses to positive 
%ones} 

%%%%%%%%%%%%%%%%%%%%%%%%%%%%%%%%%%%%%%%%%%%%%%%%%%%%%%%%%%%%%% 
\subsection{$p=4$} 
%%%%%%%%%%%%%%%%%%%%%%%%%%%%%%%%%%%%%%%%%%%%%%%%%%%%%%%%%%%%% 

\begin{figure}[bth] 
 \centerline{\psfig{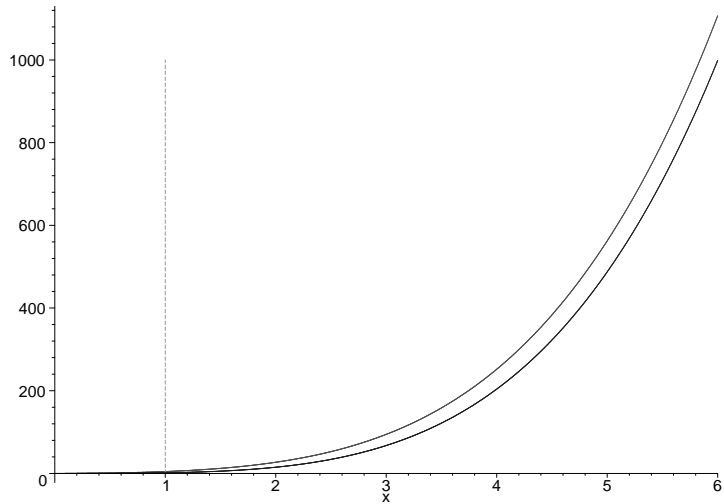}} 
 \caption{The values of the {\it masses} for D4} 
 \label{fig:d4} 
 \end{figure} 
 
The behavior of Dp branes for $p>3$ is, as explained before, very 
different. In this subsection we discuss the D4. 
 
In this particular case we are able to provide a closed-form 
solution for the \pg\ . The metric is 
 \be 
 ds^2=-4dx^+ dx^- 
-{3\m^4\over 4(E^2 -\m^4x_+^2)^2}\bigg[ 
(2E^2-\m^4x_+^2)\sum\limits_{i=1}^5 x_i^2 
 + (6E^2-\m^4x_+^2)\sum\limits_{j=6}^8 x_j^2 \bigg](dx^+)^2 
+ \sum\limits_{i=1}^8 dx_i^2, 
 \ee 
where $0\le x_+ \le {E\over \m^2}$. 
 
\section{Summary and discussion} 
An important goal of the gauge/string  duality is to develop a 
dual picture of QCD. Since this is a formidable task, much 
simpler systems that include supersymmetry and conformal 
invariance were analyzed first. At a later stage the duality 
associated with non-conformal gauge theories was also 
addressed\cite{pw}. The corresponding RG flows were mapped into 
the evolution of certain SUGRA modes. It the present work an 
attempt has been made to go a step forward by mapping the RG 
flows into an evolution on the world sheet of the string model 
derived using the Penrose\cite{penrose} limit. In fact we argued 
that an essential part of the flow is captured by a QM particle 
theory on the tdppw background. In this language the system is 
mapped into a harmonic oscillator with time dependent frequency. 
We use the tool kit that has been developed to solve this well 
known QM problem to gain insight on the operator mixing of the 
gauge theory. The main point that we have made in this regard is 
that the QM associated with the tdppw limit is {\it exactly 
solvable}. 
 
We then demonstrated these ideas by explicitly constructing the 
Penrose limit of the PW solution. We showed that in the IR fixed 
point one can take two null geodesics. The associated gauge 
theory flows were analyzed. It is remarkable that the killing 
vectors deduced from the limit on the geometry side were mapped 
into global currents of the gauge theory that led to a simple 
structure of the gauge theory operators duals of the low lying 
string states. 
 
Another class of models that we have analyzed is that of the near 
horizon limit of the large number of $D_p$ branes\cite{cobi}. We 
wrote the equations of the null geodesic lines and the masses of 
the string model. It turns out that for $p<3$ the masses--squared 
become very negative, namely there is a large upside--down 
potential unbounded from below. This signals that the world sheet 
theory has become inadequate, and points to the necessity for a 
different description of the IR region. In the SUGRA picture this 
was caused by large string coupling or large curvature. 
 
The work on this paper has raise several  related open questions 
that are under current investigation. Here is a short list of 
them: (i) We have made only a very limited use of the knowledge 
about the QM system. A natural question to ask is how to map all 
the properties of this system to those of the dual gauge theory. 
In particular one can construct special states like coherent and 
squeezed states. It is not clear to what do these state translate 
in the gauge theory operators language. (ii) For what tdppw 
string models can one provide an exact solution similar to the 
exact solution of the QM problem. (iii) The Penrose limit at any 
generic point along the (PW) flow and not only at the fixed 
points, with a better understanding for the roles of the 3--forms. 
(iv) The study of tdppw of SUGRA space-times that are duals to 
confining gauge theories. (v) The precise map of the CS equation 
in the tdppw string picture.

%%%%%%%%%%%%%%%%%%%%%%%%%%%%%%%%%%%%%%%%%%%%%%%%%%%%%%%%%%%%%%%%%%%%%%%%%%%%%%% 
 
\begin{center} 
{\large  {\bf Acknowledgments}} 
\end{center} 
We have benefited from discussions with 
D. Berenstein, T. Brun, M. Einhorn, A.Hashimoto, F. Larsen J. Maldacena, 
H. Nastase, S. Sethi, M. Strassler, C. Thorn and D. Vaman. 
We are specially grateful to J. Maldacena for a careful reading of the manuscript and useful discussions 
and to N. Seiberg for several illuminating conversations.

%%%%%%%%%%%%%%%%%%%%%%%%%%%%%%%%%%%%%%%%%%%%%%%%%%%%%%%%%%%%%%%%%%%%%%%%%%%%%% 
\appendix{Penrose limit of a general metric} 
%%%%%%%%%%%%%%%%%%%%%%%%%%%%%%%%%%%%%%%%%%%%%%%%%%%%%%%%%%%%%%%%%%%%%%%%%%%%%% 
In this appendix we provide the Penrose-G\"uven limit for the 
Pilch-Warner background for an arbitrary value or $r$. The purpose 
is to outline how we can, in principle, trace a particular 
geodesic throughout the flow, that is, for values of $r$ in both 
the IR and UV regions. Since the Pilch-Warner solution is not 
known analytically for any value of the $r$ our calculation is 
limited to the regions where the solution is known. However, it 
allows us to understand better the general scheme of Penrose 
limits in the presence of RG flows. 
 
We will divide the problem into two stages. The first one is 
general and we apply its results throughout the paper. Consider a 
metric of the form: 
\begin{equation} 
ds^2=-g_{tt}dt^2 +g_{rr}^2 dr^2 + g_{\psi\psi}^2 d\psi^2 
+g_{ij}dY^i dY^j, 
\end{equation} 
where $g_{ij}(r)$, that is, is the metric is a function of $r$ only. 
The effective Lagrangian for the null geodesic is 
\begin{equation} 
 \label{effective} 
 {\cal L}= -g_{tt}(r)\dot{t}^2 +g_{rr}(r) \dot{r}^2 + g_{\psi\psi}(r) \dot{\psi}^2, 
\end{equation} 
where dots represent derivatives with respect to the affine 
parameter $U$. Since the effective Lagrangian is explicitly 
independent of  $t$ and $\psi$ we have the following equations of 
motion 
\begin{eqnarray} 
 \label{Uonly} 
 \dot{t}&=&{E\over g_{tt}}, \qquad \dot{\psi}={\mu\over g_{\psi\psi}}, 
 \nonumber \\ 
 \dot{r}^2 &=& {E^2\over g_{rr}g_{tt}}-{\mu^2\over g_{rr}g_{\psi\psi}} 
\end{eqnarray} 
Introducing new coordinates $(U,V,\Phi)$ such that 
\begin{equation} 
r=r_U, \quad t=t_U -{v\over E}+\mu \phi, \quad \psi =\psi_U+E\Phi, 
\end{equation} 
where $x^i_U$, a function of $U$ only, is determined by 
(\ref{Uonly}) we guarantee that $g_{UU}=g_{U\Phi}=0$ and 
$g_{UV}=1$. The new metric becomes 
\begin{equation} 
ds^2=2dUdV -{1\over E^2}g_{tt}dV^2 +{2\mu\over E}g_{tt}dV d\Phi 
+(E^2g_{\psi\psi}-\mu^2g_{tt})d\Phi^2 +g_{ij}dY^i dY^j. 
\end{equation} 
Taking the Penrose limit means 
\begin{eqnarray} 
ds^2 &=& \Omega^{-2} ds^2(\Omega), \nonumber \\ 
U&=&u, \quad V=\Omega^2 v, \quad \Phi=\Omega\phi, \quad Y^i=\Omega y^i, \nonumber \\ 
\Omega&\to& 0. 
\end{eqnarray} 
The resulting metric is 
\begin{equation} 
ds^2=2dudv+(E^2g_{\psi\psi}-\mu^2g_{tt})d\phi^2 +g_{ij}dy^i dy^j, 
\end{equation} 
where the metric entries, $g_{\alpha\beta}$, are only a function 
of $u$. This metric can be taken to Brinkman coordinates. In 
particular, the mass associated $\phi$ will be 
\begin{equation} 
 m_{\phi}^2=-{\big[\sqrt{E^2g_{\psi\psi}-\mu^2g_{tt}}\big]''\over 
 \sqrt{E^2g_{\psi\psi}-\mu^2g_{tt}}}, 
\end{equation} 
where the prime means derivatives with respect to $x^+=u$. For a 
diagonal $g_{ij}$ the mass associated with the coordinate 
$x_i$ is 
\be 
m_i^2 =- {(\sqrt{g_{ii}})''\over \sqrt{g_{ii}}}. 
\ee

We now turn to the 
Pilch-Warner solution which can be written as: 
 \begin{eqnarray} 
 ds_{10}^2&=& \O^2\left(e^{2A}\big[-dt^2 +dx_1^2 +dx_2^2 +dx_3^2\big] 
 + dr^2 \right) 
 + ds_5^2, \nonumber \\ 
 ds_5^2 &=& g_{\t\t}d\t^2 + g_{\t_1\t_1}d\t_1^2+ g_{\p\p}d\p^2 
 + g_{\p_1\p_1}d\p_1^2+g_{\psi_1\psi_1}d\psi_1^2 \nonumber \\ 
 &+& 2g_{\p\p_1}d\p d\p_1 +2g_{\p\psi_1}d\p d\psi_1 
 +2g_{\p_1\psi_1}d\p_1 d\psi_1. 
 \end{eqnarray} 
where 
\begin{eqnarray} 
 g_{\t\t}&=&{L^2\,\O^2\over \rr^2 \,\cosh^2\chi}, \nonumber \\ 
 g_{\t_1\t_1}&=&{L^2\,\O^2 \,\rr^4\cos^2\t\over 4\,\bar{X}_1\,\cosh^2\chi}, \nonumber \\ 
 g_{\p\p}&=&{L^2\,\O^2\,\bar{X}_2\, \sin^2\t 
 \over \rr^2 \, \cosh\chi\, \bar{X}_1^2}, \nonumber \\ 
 g_{\p_1\p_1}&=&{L^2\,\O^2\over \rr^2\,\cosh^2\chi} 
 \bigg[{\rr^6\,\cosh\chi\, \cos^2\t\over 4\bar{X}_2} 
 +{\rr^{12}\sinh^4\chi\cos^4\t\sin^2\t\over 4\bar{X}_1^2\bar{X}_2 \cosh\chi} 
 \bigg], \nonumber \\ 
 g_{\psi_1\psi_1}&=&{L^2\,\O^2\over \rr^2\,\cosh^2\chi} 
 \bigg[{\rr^6\,\cosh\chi\, \cos^2\t\,\cos^2\t_1 \over 4\bar{X}_2} 
 +{\rr^{12}\sinh^4\chi\cos^4\t\sin^2\t\cos^2\t_1 
 \over 4\bar{X}_1^2\bar{X}_2 \cosh\chi} 
 +{\rr^6\cos^2\t\sin^2\t_1\over 4\bar{X}_1}\bigg], \nonumber \\ 
 g_{\p\p_1}&=&{L^2\,\O^2\,\rr^4\,\sinh^2\chi\,\cos^2\t\,\sin^2\t 
 \over \bar{X}_1^2\cosh^2\chi}, \nonumber \\ 
 g_{\p\psi _1}&=&{L^2\,\O^2\,\rr^4\,\sinh^2\chi\,\cos^2\t\,\sin^2\t\, \cos\t_1 
 \over \bar{X}_1^2\cosh^2\chi}, \nonumber \\ 
 g_{\p_1\psi_1}&=& {L^2\,\O^2\, \rr^4\,\cos^2\t\,\cos\t_1 
 \over 4\bar{X}_2\, \cosh^2\chi} 
 \bigg[\cosh\chi 
 +{\rr^{6}\sinh^4\chi\cos^2\t\sin^2\t\over \bar{X}_1^2\cosh\chi} 
 \bigg]. 
\end{eqnarray} 
The complex 3--form ${\cal A}_{(2)}$ is: 
 \be 
 \label{A2} 
 {\cal A}_{(2)} = e^{-i(\phi+\phi_1)}\, {L^2\over 8}\bigg[i\cos\t\ d\t - 
 {\cos^2\t\sin\t\over 1 + \sin^2\t}(d\phi + \cos\t_1\ d\psi - 
 d\psi)\bigg] \wedge (d\t_1\ + i\sin\t_1\ d\psi_1) 
 \ee 
 Note that in general $g_{ij}(r,\t,\t_1)$. Following the general 
approach we would like to find coordinates in which the above 
metric takes the form: 
 \be 
ds^2=d\b(d\l+a d\b + b_i dY^i) + C_{ij}dY^i dY^j, 
 \ee 
where $a, b_i$ and $C_{ij}$ can depend on any coordinate. This 
task involve considering a geodesic for which, in principle, all 
functions will depend nontrivially on the affine parameter. One 
is therefore, to consider an effective Lagrangian of the form: 
 \be 
 \label{PWp2} 
 {\cal L}=-g_{tt}\dot{t}^2 + g_{rr}\dot{r}^2 + g_{\t\t}\dot{\t}^2 
 + g_{\t_1\t_1}\dot{\t}_1^2 + g_{ij}\dot{x}_i \dot{x}_j, 
 \ee 
where $x_i=\{\p,\p_1,\psi_1\}$. Since the metric does not 
explicitly depends on $x_i$ we will have three integrals of 
motion that reduce the equations of motion for $x_i$ to first 
order differential equations. We are thus left with three second 
order differential equations for $\{r,\t,\t_1\}$, one of which can 
be exchanged by a first order equation which is the condition of 
the geodesic being null:  ${\cal L}=0$. With similar coordinate 
transformations to those described in the main body we bring the 
metric to the required form and subsequently perform the Penrose 
limit. 
 
A different approach that exploits the symmetry of the problem is 
to used the procedure outlined above, supplemented by the approach 
used by BMN which makes use of the specific form of the metric. 
In the BMN approach the metric automatically takes the Brinkman 
form. Namely, we will expand around a particular value of $\t$ 
and $\t_1$, at this point it does not matter which particular 
value. In the main body we naturally considered, for example, 
$\t=\pi/2$ and $\t=0$. The nature of the dependence of 
$g_{ij}(r,\t,\t_1)$ on $\t$ and $\t_1$ is such that we obtain 
$g_{ij}(r,\t,\t_1)=g_{ij}(r)\big[1+ a(r){y^2\over L^2} + 
b(r){x^2\over L^2}\big]+ {\cal O}(L^{-3})$. We have, by means of 
expanding in $\t$ and $\t_1$, reduced the problem to the one 
considered at the beginning of the  appendix. We can now consider 
a geodesic that is determined motion in $(t,r,\vartheta)$ where 
$\vartheta$ is one of the angles of $S^5$ or a combination of 
them. The only difference being that in addition to the standard 
contribution to $g_{++}$ discussed in the first part of the 
appendix due to going from Rosen to Brinkman coordinates, we now 
have a contribution due to the expansion discussed above.

Although conceptually less straightforward the described method 
to the Penrose limit is very efficient technically. In the case of 
$AdS_5\times S^5: -\cosh^2\r dt^2 + d\r^2 +\sinh^2\r d\O_3^2 + 
d\t^2 + \cos^2\ d\psi^2 + \sin^2\t\tilde{\O}_3^2$   it is 
especially simple, expanding around $\r=r/L$ and $\t=y/L$ 
 
\begin{eqnarray} 
 L^{-2}\, ds_{AdS_5}^2&=& -(1+{r^2\over L^2}+{\cal O}(L^{-4}))dt^2 
 + {dr^2\over L^2}+({r^2\over L^2}+{\cal O}(L^{-4}))d\O_3^2, \nonumber \\ 
 L^{-2}\, ds_{S^5}^2&=&(1-{y^2\over L^2}+{\cal O}(L^{-4}))d\psi^2 + 
 {d y^2\over L^2}+ 
 ({y^2\over L^2}+{\cal O}(L^{-4}))d\tilde{\O}_3^2. 
\end{eqnarray} 
The effective Lagrangian analogous to the one considered in the 
previous appendix is now 
 \be 
 {\cal L}=-\dot{t}^2 + \dot{\psi}^2, 
 \ee 
making the whole task straightforward. There is one last detail 
in taking the Penrose limit that has very important physical 
implications. In the case of the Pilch-Warner background the 
metric is not diagonal. Similar to the case of the $AdS_5\times 
T^{1,1}$  one needs to shift some of the 
coordinates by a function of $x^+$. Without shifting we have that 
following (\ref{effective}) and (\ref{Uonly}) 
 \be 
 {\pd\over \pd x^+ }= \dot{t}\pd_t + \dot{r}\pd_r 
 + \dot{\psi}\pd_\psi. 
 \ee 
However, it will generically be the case that the metric contains 
some nondiagonal terms of the form $l(x^+)dx^+ d\vartheta$. We can 
cancel those terms by simply shifting $\vartheta\to \vartheta 
+ f(x^+)$, such that 
 \be 
 \label{diagonal} 
 l(x^+)dx^+ d\vartheta + p(x^+)d\vartheta d\vartheta \to 
 p(x^+)d\vartheta d\vartheta - {l^2\over 4 p}(dx^+)^2 
 \ee 
This shift changes the masses of some of the field and more 
importantly changes the definition of $P^-$ since now 
 \be 
 {\pd\over \pd x^+ }= \dot{t}\pd_t + \dot{r}\pd_r 
 + \dot{\psi}\pd_\psi-{l\over 2p} \pd_\vartheta. 
 \ee 
It might not always be possible to diagonalize the coefficient of 
$g_{++}$ which in general is of the form $A_{ij}(x^+)x^ix^j$, in 
particular we have implicitly used the fact that both terms of 
(\ref{diagonal}) are multiplied by an overall function of say 
$y^2$ or $x^2$. In case the terms in (\ref{diagonal}) are 
multiplied by different powers of $y$ or $x$ we can not 
diagonalize the metric in this and will, in general, have 
``magnetic'' terms.

\end{document}